\documentclass[12pt]{iopart}

\usepackage[utf8]{inputenc}
\usepackage{graphicx}
\usepackage{palatino}
\usepackage{bm}
\DeclareBoldMathCommand{\bfeta}{\eta}
\DeclareBoldMathCommand{\bfnabla}{\nabla}
\DeclareBoldMathCommand{\tensG}{{\sf G}}

\newcommand{\figpath}{}

\usepackage[usenames,dvipsnames,pdftex]{color}

\begin{document} 

\title{Friction forces on atoms after acceleration}
\author{
Francesco Intravaia$^{1}$,
Vanik E. Mkrtchian$^{2,6}$
Stefan Buhmann$^3$,
Stefan Scheel$^{4}$, 
Diego A. R. Dalvit$^{5}$,
and Carsten Henkel$^{2}$}
\address{$^{1}$\ Max-Born-Institut, 12489 Berlin, Germany} 
\address{$^{2}$\ Institute of Physics and Astronomy, University of Potsdam, 
Karl-Liebknecht-Str. 24/25, 14476 Potsdam, Germany}
\address{$^{3}$\ Physikalisches Institut, Albert-Ludwigs-Universit\"at Freiburg, Hermann-Herder-Str. 3, 79104 Freiburg, Germany}
\address{$^{4}$\ Institut f\"ur Physik, Universit\"at Rostock, Universit\"atsplatz 3, 18055 Rostock, Germany}
\address{$^{5}$\ Theoretical Division, MS B213, Los Alamos National Laboratory, Los Alamos, New Mexico 87545, USA}
\address{$^{6}$\ Institute for Physical Research,
Armenian Academy of Sciences, Armenia}

%%%%%%%%%%%%%%%%%%%%%%

\begin{abstract}

The aim of this paper is to revisit the calculation of atom-surface quantum 
friction in the quantum field theory formulation put forward by
Barton [\emph{New J. Phys.} {\bf 12} (2010) 113045]. 
We show that the power dissipated into field
excitations and the associated friction force depend on how
the atom is boosted from being initially at rest to a configuration 
in which it is moving at constant velocity ($v$) parallel to the planar interface.
In addition,  we point out that there is a subtle cancellation between the one-photon 
and part of the two-photon dissipating power, resulting in a leading order contribution
to the frictional power which goes as $v^{4}$.
These results are also confirmed by an alternative calculation of the average radiation force, which scales as $v^{3}$.
\end{abstract}

%%%%%%%%%%%%%%%


\submitto{\JPCM}

\section{Introduction}

The interaction of moving objects with light has been in the
focus of physics even before Einstein's \emph{annus mirabilis}
and his fundamental papers about special relativity. A seminal
contribution in this context is 
Einstein's derivation of Planck's blackbody radiation 
law~\cite{Einstein1917,Kleppner2005}: it has brought upon us 
not only the concepts of spontaneous and stimulated emission. Also the momentum
exchange between atoms and photons, and the corresponding friction
and diffusion have been shown to provide the physical picture for 
the thermalization of the velocity distribution of an atomic gas,
decades before the advent of laser cooling 
techniques~\cite{MetcalfBook}. Radiative
friction (without lasers) 
is the process where a moving atom comes to rest in the
preferred frame set by the blackbody radiation 
field~\cite{Mkrtchian2003}. (Motion
relative to the frame of the cosmic microwave background, for example,
can indeed be detected by the anisotropy in the apparent 
temperature~\cite{Lineweaver1996}.)
Quantum friction is the theorists' variant of this problem, when
the temperature is set to zero. Velocity-dependent (or drag) forces 
only appear when true relative motion is defined by the presence of
another object. In this paper,
we consider the simple case of an atom (or molecule) near a macroscopic half-space filled 
with metallic material. The distance between the atom
and the metal surface is also macroscopic (at least a few nm)
in the sense that electronic overlap is negligible. 
In this regime of distances, it is valid to use a local approximation for the optical response of the surface, its permittivity depending only on frequency.

It is instructive to draft a short summary of the long series of works dealing with the 
problem of quantum friction on an atom moving at constant velocity 
parallel to the vacuum-metal interface.
We will restrict ourselves to works that mainly used the local approximation for
the optical
response, i.e.\ those that considered macroscopic distances in the sense defined above.
It is interesting to note that various authors obtained quite different 
results for this drag force, differing both in their dependence on 
velocity and with atom-surface separation.
Unfortunately, most of these works do not critically discuss the 
others nor attempt to clarify the origins of the differences. 
One of the earliest works on the problem was undertaken by 
Mahanty~\cite{Mahanty1980}, who computed the velocity dependence of 
the drag force on a moving molecule. It was found that the 
quantum friction force scales as $v z^{-5}$ for small velocity
$v$ and large separation $z$ between molecule and surface.
However, this calculation was criticized by various authors since 
it predicts a non-zero quantum friction even for a perfectly reflecting 
surface (which lacks the possibility of referencing relative motion, 
indeed). Another series of papers also obtained a linear dependence of 
quantum friction on velocity. Schaich and Harris~\cite{Schaich1981} 
computed dynamic corrections to van der Waals potentials for a 
neutral molecule moving above a metallic plate, and modeled the molecule 
as a dipole oscillating normal to the surface. The resulting friction
force is again linear in velocity but with a different asymptotic 
large distance dependence as $z^{-10}$. More recently, Scheel and 
Buhmann~\cite{Scheel2009} have considered a multi-level atom moving at 
constant velocity, and employed a master equation approach to solve 
for the atom dynamics in the Markov approximation. They again found a linear 
dependence on velocity and a $z^{-8}$ scaling in the near-field. These same
scalings (with slightly different numerical pre-factors) were obtained by 
Barton~\cite{Barton2010b} in a harmonic oscillator model for the atom, where 
the friction force is computed in time-dependent perturbation theory from the 
power dissipated into pairs of plasmons. 
H{\o}ye and Brevik have put forward an approach to quantum friction very similar to Barton's, and used it to compute the friction force between two atoms 
\cite{Hoye2011b} or two plates \cite{Hoye2014} and compared it to Barton's results for these particular systems.

In contrast to all the above works in the literature, various other authors 
have obtained a vanishing contribution to the atom-surface friction force linear in 
velocity. For example, Tomassone and Widom~\cite{Tomassone1997} computed the 
finite temperature friction force on molecules moving near metals using the 
image charge approach, and obtained a vanishing linear-in-$v$ quantum friction 
in the limit of zero temperature. The same conclusion was reached by Volokitin 
and Persson~\cite{Volokitin2002}, who employed fluctuation electrodynamics to 
compute the Lorentz force on a moving dipole, by Dedkov
and Kyasov~\cite{Dedkov2002}, who used the equilibrium 
fluctuation-dissipation theorem to evaluate the dipole and field correlation 
functions, and by Golyk, Kr\"uger and Kardar \cite{Golyk2013}, who evaluated the force using 
linear response relations in fluctuation electrodynamics. 
Another series of papers confirmed these results, and derived the 
first non-vanishing contribution to the quantum friction force
that scales as $v^3$. These include the works of Dedkov and 
Kyasov~\cite{Dedkov2012}, who extended their previous calculations to capture 
the nonlinear dependence of quantum friction on velocity, Pieplow and 
Henkel~\cite{Pieplow2013}, who used equilibrium fluctuation electrodynamics 
to derive a relativistically covariant formulation for the friction force, and 
Intravaia, Behunin and Dalvit~\cite{Intravaia2014}, who 
calculated the atom-surface drag force by generalizing 
fluctuation-dissipation relations to the non-equilibrium stationary
state defined by a constant velocity.

One of the goals of this paper is to revisit the calculation of quantum 
friction in probably one of the simplest and cleanest formulations of the 
problem put forward by Barton~\cite{Barton2010b}.
Within this approach, the zero-temperature friction force is computed by 
solving the joint atom+field/matter dynamics in time-dependent perturbation 
theory,
starting from an initial state in which the atom and the 
field/matter subsystems are both in their (`bare') ground states. 
As emphasized by Barton, this perturbation theory has no need to rely on
assumptions related to correlation times, linear response, or local 
thermodynamic equilibrium
which are implicit in many calculations performed with the toolbox of
fluctuation electrodynamics. One also does not require fluctuation-dissipation
relations. The challenge of this approach is that the dissipation in the
atomic system is \emph{purely radiative} and is generated self-consistently
in the perturbation series. This is in sharp contrast to a field theory
like the one reported by Volokitin and Persson~\cite{Volokitin2006}
where the basic two-point functions for atomic variables are constructed
by a re-summation procedure including radiative damping.
For simplicity, we will restrict ourselves
to the near-field regime, where quantum friction is expected to be enhanced. 
We demonstrate in particular that the power dissipated into field
excitations and the associated friction force depends on how
the atom is boosted from being initially at rest to a configuration 
in which it is moving at constant velocity parallel to the planar
interface.

The paper is organized as follows. Sec.\ref{s:model} reviews the
building blocks of the quantum field theory for the atom-field
interaction 
and gives the time-dependent state including amplitudes for
one- and two-photon processes.
In Sec.\ref{s:Barton-PA-1} we use these results to calculate
the frictional power and force in the case where the velocity of the particle
is constant for all times. 
Although this obviously requires an external energy supply to compensate
for the frictional loss, the description is actually simpler, and one
recovers
some of the results presented in Ref.\cite{Barton2010b}. 
It is shown in particular 
that the ${\cal O}( v^4 )$ contribution to the power of two-photon emission found in 
Ref.\cite{Barton2010b} (called there $P_{A}$) can be explained in terms of
this special trajectory. Sec.\ref{s:Barton-PA-1} also provides an
alternative picture where the expectation value of the force
operator is computed in the time-dependent state. Its stationary
value at long times is found to scale with the velocity like $\sim v^3$.
Sec.\ref{s:generic-path} contains the main results of this paper. 
The calculation of the radiated power is generalized 
to more realistic trajectories where the atom starts
at rest and is accelerated to a constant final velocity.
We discuss the role of the finite
duration of the acceleration and show that: (i) the results presented 
in Ref.\cite{Barton2010b} depend of the specific choice of the atom's trajectory; 
(ii) that the power bookkeeping in Ref.\cite{Barton2010b} is incomplete and needs 
to be complemented with the power needed to create the excited state. If this
is done, we again find a frictional force that scales as $\sim v^3$ with 
the velocity.
Sec.\ref{s:fluctQED} provides a review of two 
approaches~\cite{Scheel2009,Intravaia2014} that describe quantum
friction within the framework of fluctuation electrodynamics.
Some technical material is relegated to the appendices.

%%%%%%%%

\section{The model}
\label{s:model}

\begin{figure}[htbp] %  figure placement: here, top, bottom, or page
   \centering
   \includegraphics[height=40mm]{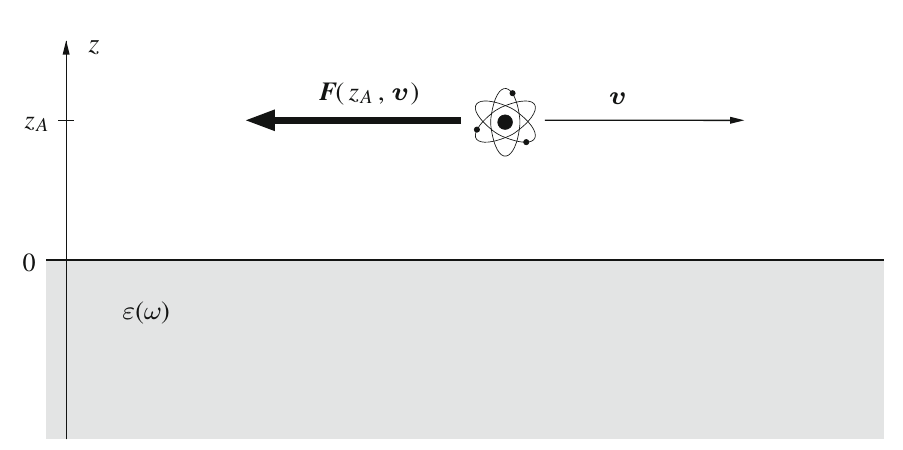} 
   \hspace*{2.5mm}
   \includegraphics[height=40mm]{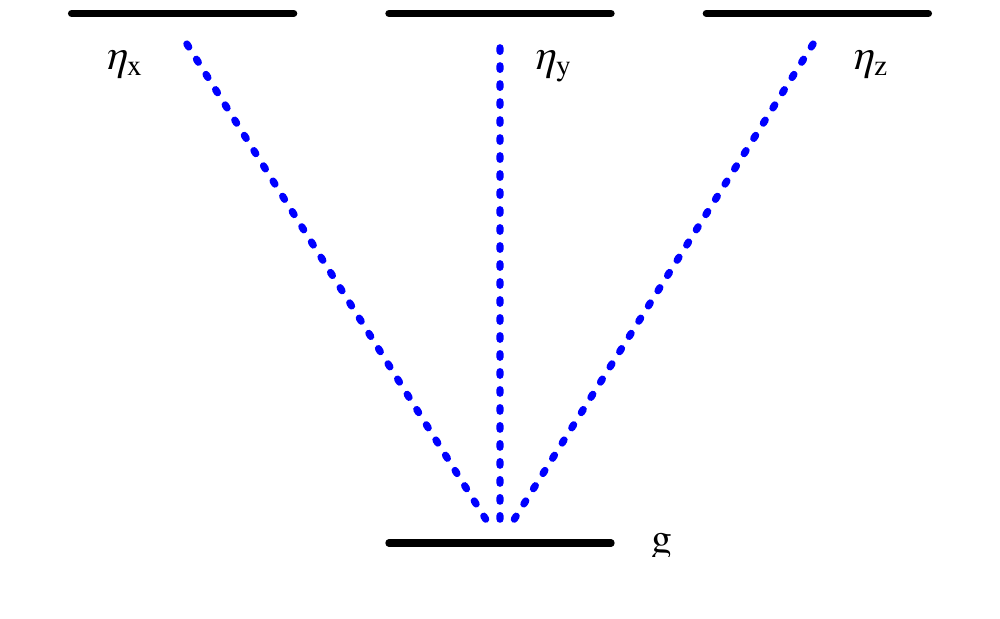} 
   \caption[]{(\emph{left}) Sketch of the considered geometry.
   (\emph{right}) Atomic energy levels.}
   \label{fig:sketch-setup}
\end{figure}

Our discussion is based on Refs.\cite{Barton2010b,Barton1997} 
where one of the simplest field theories for atom-photon interactions
is developed. The physical situation is sketched in 
Fig.\ref{fig:sketch-setup} (left): a point-like atom moves at constant
velocity ${\bf v}$ parallel to a half-space that responds linearly
to the electromagnetic field. The distance $z = z_A$ of the atom,
kept fixed, is taken much smaller than the relevant wavelengths
(non-retarded regime) so that the field can be described by an
electric potential $\Phi$
[Eq.(\ref{eq:electric-potential}) below]. 
The half-space is absorbing light,
broadening the surface plasmon resonance. For simplicity, we
still call `photons' the elementary excitations of the field,
although `plasmon-polariton' or `medium-assisted polariton' 
would be more appropriate names. The atom is described by a few
low-lying states (Fig.\ref{fig:sketch-setup}, right), and its position
follows a prescribed trajectory ${\bf r}( t )$.
Our goal is to calculate the radiative friction force ${\bf F}$
and the frictional power $P = - \dot {\bf r} \cdot {\bf F}$ that must be 
supplied by the external agency that keeps the atom on its path.

%%%%%%%%%

\subsection{Relevant states and observables}

The lowest quantum states of the atom are taken by analogy to the
${\rm 1s}$ and ${\rm 2p}$ level of the hydrogen atom:
they are denoted
$| g \rangle$ for the 1s state, 
the three degenerate ${\rm 2p}_i$ states are written 
$| \vec{\eta} \rangle$. The unit vector $\vec{\eta}$ is taken 
from a set $\{ \vec{\eta} \}$ forming an orthonormal basis
that we may assume real without loss of generality.
The Bohr transition frequency between the levels is $\Omega$. We focus in this paper 
on transitions among these energy levels only 
and mention briefly where additional states would appear.
The nonzero matrix elements of the electric dipole operator 
$\vec{\hat{D}}( t )$ in the interaction picture 
are 
\begin{equation}
\langle g | \vec{\hat{D}}( t ) | \vec{\eta} \rangle = 
\vec{\eta} \, d \, {\rm e}^{ - {\rm i} \Omega t}
\,,\qquad
\langle \vec{\eta} | \vec{\hat{D}}( t ) | g \rangle = 
\vec{\eta} \, d \, {\rm e}^{ {\rm i} \Omega t}~,
	\label{eq:dipole-matrix-elements}
\end{equation}
where the transition dipole matrix element $d$ is the basic coupling 
constant of the field theory. 
It determines, for example, the static polarizability
$\alpha = (2 d^2 )/(\hbar \Omega)$
[Eq.(2.2) of Ref.\cite{Barton2010b}].
Using rotational symmetry, one also has the identity 
$\sum_{\vec{\eta}} \eta_i \eta_j = \delta_{ij}$
when the three excited states are summed over.

%%%%%%%

The atom+field coupling (summation over double indices is 
assumed hereafter)
\begin{equation}
\hat{V}( t ) = - \hat{D}_i( t ) \hat{E}_i( \vec{r}( t ), t )
=
\hat{D}_i \partial_i \hat{\Phi}( \vec{r}( t ), t )
	\label{eq:dE-coupling}
\end{equation}
is explicitly time-dependent via the atomic trajectory
$\vec{r}( t )$. We will also often use the notation ${\bf r}( t )$ for 
the path in the $xy$-plane parallel to the surface placed at $z=0$. 
In our approach the force
acting on the atom parallel to this plane is given by the 
operator~\cite{Jackson}
\begin{equation}
\hat{{\bf F}}( t ) = 
\hat{D}_i( t ) 
\bfnabla \hat{E}_i( \vec{r}( t ), t ).
	\label{eq:force-operator}
\end{equation}
More general situations would include 
higher-order multipole moments of 
the atomic charge and current distribution (magnetic dipole,
electric quadrupole \ldots) and the time derivative of the 
electromagnetic momentum $\vec{d} \times \vec{B}$. The latter
includes 
the so-called R\"ontgen interaction that takes into account
the transformation of the electromagnetic field into the frame co-moving
with the atom. This interaction is relevant at larger (retarded) 
distances~\cite{Scheel2009}.

The field operator is expanded in a plane-wave basis of 
elementary excitations (photons) and evolves freely according to
\begin{equation}
\hat{\Phi}( \vec{r}, t )
= \int\!{\rm d}^2k\!\int\limits_{0}^{\infty}\!{\rm d}\omega
\left( \hat{a}_{{\bf k}\omega}\,
\phi_{{\bf k}\omega}
\exp( {\rm i} {\bf k} \cdot {\bf r}
- {\rm i} \omega t )
 + \mbox{h.c.} \right)~.
	\label{eq:electric-potential}
\end{equation}
The bosonic operators satisfy the commutation relation $[\hat{a}^{\phantom\dag}_{{\bf k}\omega}, \hat{a}^\dag_{{\bf k}'\omega'}]=\delta( {\bf k} - {\bf k}' ) \delta( \omega - \omega' )$
and the `one-photon amplitudes' are given by
[Eq.(2.4) of \cite{Barton2010b}]
\begin{equation}
\phi_{{\bf k}\omega}
= 
\frac{ \sqrt{ \omega \Gamma \omega_p^2 / 2 } }{ 
\omega^2 + {\rm i} \omega \Gamma - \omega_S^2 
}
\sqrt{ \frac{ \hbar }{ 2\pi^2 k } }
\, {\rm e}^{ - k z }~,
	\label{eq:def-coupling-phi-kappa}
\label{eq:}
\end{equation}
where the frequencies $\omega_p$, $\omega_S$, and $\Gamma$ parametrize
the dielectric function $\varepsilon( \omega )$ of the half-space.
We note in particular the relation
\begin{equation}
|\phi_{{\bf k}\omega}|^2 = 
\frac{ \hbar }{ 2\pi^2 }
\frac{ {\rm e}^{ - 2 k z } }{ k }
\mathop{\rm Im}R ( \omega )
=
\frac{ \hbar }{ 2\pi^2 }
\frac{ {\rm e}^{ - 2 k z } }{ k }
\mathop{\rm Im}
\frac{ \varepsilon( \omega ) - 1 }{ \varepsilon( \omega ) + 1 }~,
\label{eq:coupling-and-ImR}
\end{equation}
where $R( \omega )$ is the non-retarded reflection coefficient
of the surface. The frequency $\omega_S$ gives the surface plasmon
resonance and $\Gamma$ its broadening (complex pole of $R( \omega )$). 
With this expansion for the photon field, the force 
operator~(\ref{eq:force-operator}),
for example, takes the form
\begin{eqnarray}
\hat{{\bf F}}( t ) &=&
\int\!{\rm d}^2k
\int\limits_{0}^{\infty}\!{\rm d}\omega\,
\left(
{\bf k}
(\vec{\hat{D}}( t )
\cdot
\vec{ k }) \phi_{{\bf k}\omega} 
\hat{a}_{{\bf k}\omega}  %\, 
\exp( {\rm i}{\bf k} \cdot {\bf r}(t) - {\rm i} \omega t )
+
\mbox{h.c.}
\right)~,
\label{eq:force-operator-expansion}
\end{eqnarray}
where the three-dimensional wave vector $\vec{k} =
( {\bf k}, {\rm i} k )$ is in fact complex with $\vec{k} \cdot
\vec{k} = 0$ and $\vec{k} \cdot \vec{k}^* = 2 k^2$. Note that
this is a `skew' operator that connects quantum states with
different photon numbers (for the field) and different energy
levels (for the atom).

Since part of our focus will be on the power radiated into photons
and pairs of photons, let us introduce
\begin{eqnarray}
P_1 &=&
\lim_{t\to\infty}
\sum_{\vec{\eta}} \int\!{\rm d}^3\kappa\,
\hbar( \Omega + \omega )
\frac{ | \langle \vec{\eta}, \kappa | \Psi( t ) \rangle |^2 }{ t }
\label{eq:def-excitation-power}\\
P_2 &=& 
\frac12 \lim_{t\to\infty}
\int\!{\rm d}^3\kappa_1\int\!{\rm d}^3\kappa_2\,
\hbar( \omega_1 + \omega_2 )
\frac{ | \langle g, \kappa_1 \kappa_2 | \Psi( t ) \rangle |^2 }{ t }~,
\label{eq:def-two-photon-power}
\end{eqnarray}
where
$| \langle \vec{\eta}, {\bf k}\omega | \Psi( t ) \rangle |^2$ and $| \langle g,
{\bf k}_1\omega_1 {\bf k}_2\omega_2 | \Psi( t ) \rangle |^2$  are the
probabilities
of finding the atom in an excited state and one and 
and the two-photon, respectively
(see Secs.\ref{s:atom-field-states} and \ref{s:generic-path}). 
We also used the compact label $\kappa = {\bf k}\omega$ and the factor
$\frac12$ accounts for double counting the
symmetric two-photon states. The long-time limit is to be understood
within time-dependent perturbation theory: $t$ is typically not longer
than a fraction of the relevant life times.
The two-photon power has been calculated
in Ref.\cite{Barton2010b}; we review the evaluation of the integrals
in \ref{a:power-Barton}. This calculation is generalized in 
Sec.\ref{s:generic-path}, where also a partial cancellation between
$P_1$ and $P_2$ is found.

%%%%%%%%%%

\subsection{Atom+field states}
\label{s:atom-field-states}

\begin{figure}[tbhp] %  figure placement: here, top, bottom, or page
   \centering
   \includegraphics*[height=45mm]{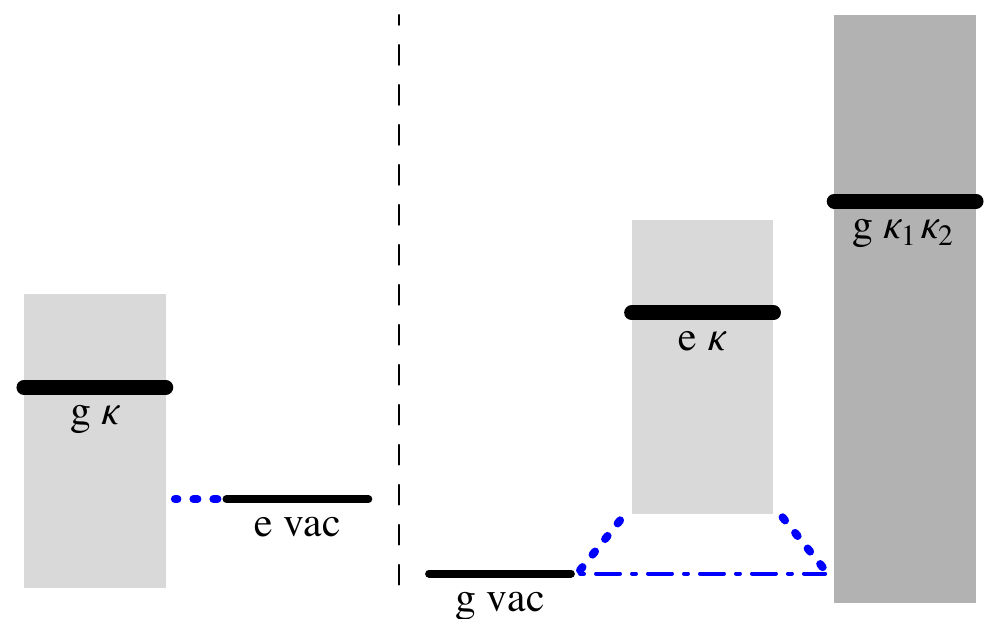} 
   \caption[]{(Color online) Schematic overview of the relevant states for the
   atom+field theory. The atomic states are labelled
   ${\rm g}$ and ${\rm e}$ (actually, ${\rm e}$ is threefold
   degenerate). The notation $\kappa = {\bf k}\omega$
   collects the quantum numbers of the field states, ${\rm vac}$ is the
   vacuum state.
   Two-photon
   states are denoted $\kappa_1\kappa_2$. To the left of the vertical
   dashed line are states that are not directly coupled to the
   ground state by the
   atom+field interaction, illustrated by the dotted blue lines.
   The thick black lines indicate the position
   of the surface plasmon resonance $\omega_S$, the shaded grey area
   illustrates its broadening (not to scale) over a range 
   $\Gamma$ (intrinsic) and due to the Doppler shift 
   ${\bf k}\cdot {\bf v}$. The Doppler shift is responsible for the
   appearance of negative frequencies, intuitively understood as
   measured in the frame co-moving with the atom. The thin dash-dotted
   line illustrates the two-photon process $g {\rm vac} 
   \leftrightarrow   g \kappa_1 \kappa_2$.
   }
   \label{fig:level-scheme}
\end{figure}

\noindent
In our perturbative approach the initial state is given by the tensor 
product of 
atomic ground state and zero photons,
$| g, {\rm vac} \rangle$, while the interaction is represented via the
operator $\hat{V}(t)$.
An expansion up to the third order in the coupling constant $d$ of the atom+field state $| \Psi( t ) \rangle$ leads to
\begin{eqnarray}
| \Psi( t ) \rangle =
(1 + c_0^{(2)}( t )) | g, {\rm vac} \rangle
%\nonumber\\
&+& 
\sum_{\vec{\eta}} \!\int\!{\rm d}^3\kappa\,
(c_1^{(1)}( t ) + c_1^{(3)}( t )) | \vec{\eta}, \kappa \rangle
\nonumber\\
&+&
\frac12 \!\int\!{\rm d}^3\kappa_1 \,{\rm d}^3\kappa_2\,
c_2^{(2)}( t ) | g, \kappa_1 \kappa_2 \rangle
+ \ldots
	\label{eq:state-expansion-FD}
\end{eqnarray}
where $c_n^{(p)}( t )$ denotes the transition amplitudes for states with $n$ photons in
the $p$'th perturbative order and can be obtained by using the standard techniques of
perturbation theory. The relevant matrix elements are given by
\begin{eqnarray}
\langle g, {\rm vac} | \hat{V}( t ) | \vec{\eta}, \kappa \rangle
&=&
{\rm i} \,d \,(\vec{\eta} \cdot \vec{k}) \phi_{\kappa}
\exp[ -{\rm i}( \Omega + \omega )t + {\rm i} {\bf k} \cdot {\bf r}( t )]~,
	\label{eq:ge-element-for-F}
\\
\langle \vec{\eta}, \kappa | \hat{V}( t ) | g, \kappa_1 \kappa_2 \rangle
&=&
{\rm i} \,d \,(\vec{\eta} \cdot \vec{k}_1) \phi_{\kappa_1}
\,{\rm e}^{ {\rm i}( \Omega - \omega_1 )t + {\rm i} {\bf k}_1 \cdot {\bf r}( t )}
\delta( \kappa - \kappa_2 )~
\nonumber\\
&& {} +
{\rm i} \,d \,(\vec{\eta} \cdot \vec{k}_2) \phi_{\kappa_2}
\,{\rm e}^{ {\rm i}( \Omega - \omega_2 )t + {\rm i} {\bf k}_2 \cdot {\bf r}( t )}
\delta( \kappa - \kappa_1 )~,
	\label{eq:eg2-element-for-F}
\\
\langle \vec{\eta}, {\rm vac} | \hat{V}( t ) | g, \kappa \rangle
&=&
{\rm i}\,d \,(\vec{\eta} \cdot \vec{k}) \phi_{\kappa}
\exp[ {\rm i}( \Omega - \omega )t + {\rm i} {\bf k} \cdot {\bf r}( t )]
	\label{eq:eg-element-for-F}~.
\end{eqnarray}
It is important to note that the matrix elements depend on the detail of the path ${\bf r}( t )$. 
% CH
Let consider first the simple case of a constant velocity, i.e. 
${\bf r}( t ) = {\bf v} t$. (Corrections arising from a realistic trajectory 
including an acceleration stage (`launch') are discussed in 
Sec.\ref{s:generic-path}.)
In this case time-dependent perturbation theory leads to
\begin{eqnarray}
c_1^{(1)}( t )
&=&
\frac{ {\rm i} \, d \, (\vec{\eta} \cdot \vec{k})^* \phi_{\kappa}^* }{ 
\hbar ( \Omega + \omega' - {\rm i} \lambda) }
\exp[ {\rm i}( \Omega + \omega' ) t ]
	\label{eq:c1-amplitude}
\\
c_2^{(2)}( t ) &=& 
-
\frac{ d^2 (\vec{k}_1 \cdot \vec{k}_2)^* 
	\phi_{\kappa_1}^* \phi_{\kappa_2}^*
\exp[ {\rm i}( \omega_1' + \omega_2' )t ]
}{ 
\hbar^2 ( \omega_1' + \omega_2' - {\rm i} \lambda ) }
\nonumber\\
&& \qquad \times \bigg\{
\frac{ 1 }{ \Omega + \omega_1' - {\rm i} \lambda }
+
\{ 1 \leftrightarrow 2 \}
\bigg\}
% + \ldots
%
	\label{eq:c2-amplitude}
\end{eqnarray}
where the positive infinitesimal $\lambda$ ensures that the atom-field
interaction is adiabatically switched on in the past. 
The scalar product $(\vec{ k }_1 \cdot \vec{ k }_2)^*
= {\bf k}_1 \cdot {\bf k}_2 - k_1 k_2$ arises from
the summation over the three excited states $| \vec{\eta} \rangle$ and we have used the shorthand $\omega' = \omega - {\bf k} \cdot {\bf v}$
for the Doppler-shifted frequency (as `seen' by the moving atom).

The second-order correction $c_0^{(2)}( t )$ needs a special handling
because it involves the energy shift of the
state $| g, {\rm vac} \rangle$ and the rate for the process
$| g, {\rm vac} \rangle \to | \vec{\eta}, \kappa \rangle$:
\begin{equation}
c_0^{(2)}( t ) = 
{\rm i} \sum_{\vec{\eta}}\int\!{\rm d}^3\kappa\,
\frac{ d^2 \, |\vec{\eta} \cdot \vec{k}|^2 |\phi_{\kappa}|^2 }{ 
\hbar^2 ( \Omega + \omega - {\bf k} \cdot {\bf v} - {\rm i} \lambda) }
\int\limits_{-\infty}^{t}\!{\rm d}t'\, {\rm e}^{ \lambda t' }~.
	\label{eq:c0-amplitude-2}
\end{equation}
The second integral integral formally evaluates to $\lambda^{-1}
{\rm e}^{ \lambda t} \approx t + \lambda^{-1}$ for $\lambda\to 0$ so that 
we identify the frequency shift and inverse lifetime from
$1 + c_0^{(2)}( t ) \approx \exp( - {\rm i} \delta E_g t / \hbar - 
\gamma_g t / 2)$. This yields
\begin{equation}
- {\rm i} \delta E_g - \frac{ \hbar \gamma_g }{ 2 } =
{\rm i} \sum_{\vec{\eta}}\int\!{\rm d}^3\kappa\,
\frac{  d^2 \, |\vec{\eta} \cdot \vec{k}|^2 |\phi_{\kappa}|^2 }{ 
\hbar ( \Omega + \omega - {\bf k} \cdot {\bf v} - {\rm i} \lambda) }~.
	\label{eq:0-level-shift-and-rate}
\end{equation}
The level shift $\delta E_g$ arising from Eq.(\ref{eq:0-level-shift-and-rate})
has been worked out by Barton, Eq.(3.2) of Ref.\cite{Barton2010b}.
For the lifetime, we get the `Golden Rule' result \cite{Boussiakou2002}
\begin{equation}
\gamma_g = \frac{ 2 \pi }{ \hbar^2 }
\sum_{\vec{\eta}}\int\!{\rm d}^3\kappa\,
| \langle \vec{\eta}, \kappa | \hat{V} | g, {\rm vac} \rangle|^2
\delta( \Omega + \omega - {\bf k} \cdot {\bf v} )~,
	\label{eq:Golden-Rule-lifetime}
\end{equation}
where the atomic motion leads to the Doppler shift of the final
photon frequency. This integral is, however, exponentially small
for reasonable parameters, as can be seen as follows. 
The sum over the excited states gives 
$\sum_{\vec{\eta}} |\vec{\eta} \cdot \vec{k}|^2 =
\vec{k} \cdot \vec{k}^* = 2 k^2$. Inserting the coupling 
strength $|\phi_{\kappa}|^2$ from Eq.(\ref{eq:coupling-and-ImR}), 
we get
\begin{equation}
\gamma_g = 
\frac{ \alpha \Omega }{ \pi }
\int\!{\rm d}^3\kappa\, k
\, {\rm e}^{ - 2 k z }
\mathop{\rm Im}R( \omega )
\delta( \Omega + \omega - {\bf k} \cdot {\bf v} )~.
	\label{eq:gamma-g-exp-small}
\end{equation}
The resonance condition
$0 = \Omega + \omega - {\bf k} \cdot {\bf v}$
jointly with $\omega \ge 0$
limits the domain for the $k$-integration to 
${\bf k} \cdot {\bf v} \ge \Omega$. 
We then have $k \ge \Omega / v$,
and the exponential gives a scaling of this integral 
proportional to ${\rm e}^{ - 2 \Omega z / v }$. We follow here the
same strategy as Ref.\cite{Barton2010b} and neglect contributions
that show such an exponential scaling, assuming that the 
velocity is small enough: $v \ll \Omega z$. [For a lithium beam
at $10\,{\rm keV}$ and $10\,{\rm nm}$ distance, $v / \Omega z 
\approx 0.02$.] 
The physical interpretation of this process is
the following [see also Ref.\cite{Pieplow2014} in this issue]: 
the atomic motion leads to an anomalous
Doppler shift ($\omega' < 0$ in the co-moving frame,
while $\omega > 0$) that makes the `spontaneous excitation' of 
the ground state possible, similar to Cherenkov 
radiation~\cite{Ginzburg1996,Maghrebi2013b}.
The rate for this process is, however, extremely slow because of
the threshold set by the atomic Bohr frequency, 
$\omega' = - \Omega$.

For the third-order correction
to the one-photon process one gets:
\begin{eqnarray}
c_1^{(3)}( t ) &=& 
- \frac{ {\rm i} }{ 2 \hbar }
\int\!{\rm d}^3\kappa_1\,{\rm d}^3\kappa_2
\int\limits_{-\infty}^{t}\!{\rm d}t' \,
\langle \vec{\eta}, \kappa | \hat{V}( t' ) | g, \kappa_1 \kappa_2 \rangle
% \langle g, \kappa_1 \kappa_2 | \Psi( t' ) \rangle
c^{(2)}_2( t' )
	\nonumber
	\\
&& 
- \frac{ {\rm i} }{ \hbar }
\int\limits_{-\infty}^{t}\!{\rm d}t' \,
\langle \vec{\eta}, \kappa | \hat{V}( t' ) | g, {\rm vac} \rangle
%\langle g, {\rm vac} | \Psi( t ) \rangle^{(2)}~.
c^{(2)}_0( t' )~.
	\label{eq:c3-perturb-iteration}
\end{eqnarray}
Due to the Bose symmetry, the two
terms in the matrix element~(\ref{eq:eg2-element-for-F}) give
the same contribution to the first line of (\ref{eq:c3-perturb-iteration}), and we get
\begin{eqnarray}
c_1^{(3)}( t ) &=& 
\frac{ {\rm i}\, d^3 \phi_{\kappa}^* 
	\,{\rm e}^{ {\rm i} ( \Omega + \omega' ) t }
	}{ 
	\hbar^3 ( \Omega + \omega' - {\rm i} \lambda ) 
	}
\int\!{\rm d}^3\kappa_1\,
\frac{ (\vec{\eta} \cdot \vec{k}_1) |\phi_{\kappa_1}|^2 (\vec{k}_1 \cdot \vec{k})^* 
}{ 
( \omega_1' + \omega' - {\rm i} \lambda ) 
}
\nonumber\\
&& \quad \times
\bigg\{
\frac{ 1 }{ \Omega + \omega_1' - {\rm i} \lambda }
+
\frac{ 1 }{ \Omega + \omega' - {\rm i} \lambda }
\bigg\}
\nonumber\\
&& 
- \frac{ {\rm i} \, d ( \vec{\eta} \cdot \vec{k} )^* \phi_\kappa^*
\,{\rm e}^{ {\rm i} ( \Omega + \omega' ) t }
	}{ \hbar 
( \Omega + \omega' - {\rm i} \lambda ) }
\Big( \frac{ \gamma_g }{ 2 } + \frac{ {\rm i} \delta E_g }{ \hbar } \Big)
\Big( t + \frac{ {\rm i} }{ \Omega + \omega' - {\rm i} \lambda } \Big)~,
	\label{eq:c3-adiabatically}
\end{eqnarray}
where the last line features again a linearly increasing part.
This amplitude will be used in 
Sec.\ref{s:ave-force} to calculate the average force 
operator for an atom in constant motion. A correction 
$\delta c_1^{(3)}( t )$ arising from the acceleration stage is 
discussed in Sec.\ref{s:tentative-interpretation} and related
to the energy stored in the excited state 
$| \vec{\eta}, \kappa \rangle$.

%%%%%%%

\section{Frictional power and force for constant velocity}
\label{s:Barton-PA-1}

% CH
The two-photon power $P_2$ [Eq.(\ref{eq:def-two-photon-power})]
has been introduced and evaluated in detail in Ref.\cite{Barton2010b}. 
There, the calculation was performed for a trajectory where the atom is
at rest for $t<0$ and having a constant velocity ${\bf v}$ for $t>0$. 
We analyze the corresponding process
% from Ref.\cite{Barton2010b} 
in the following 
Sec.\ref{s:generic-path}, and review an alternative calculation reported in 
Ref.\cite{Scheel2009} in Sec.\ref{s:macro-QED}. In this section, we evaluate
$P_2$ in the case of constant velocity.

\subsection{One- and two-photon emission}

We find
that the only relevant amplitude $c_2^{(2)}( t ) 
= \langle g, \kappa_1\kappa_2 | \Psi( t ) \rangle$
{}[Eq.(\ref{eq:c2-amplitude})]
translates into the following differential emission rate
\begin{eqnarray}
\hspace*{-10mm}
{\rm d} w_2 &=& \lim_{t\to\infty} \frac{ |\langle g, \kappa_1\kappa_2 
	| \Psi( t ) \rangle|^2 }{ t }
	\frac{ {\rm d}^3\kappa_1\,{\rm d}^3\kappa_2 }{ 2 } 
\nonumber\\
&=&
\frac{ d^4 |\vec{k}_1 \cdot \vec{k}_2|^2 
	|\phi_{\kappa_1}|^2 |\phi_{\kappa_2}|^2
}{ \hbar^4 } 
\bigg|
\frac{ 1 }{ \Omega + \omega_1' }
+
\{ 1 \leftrightarrow 2 \}
%\frac{ 1 }{ \Omega + \omega_2' - {\rm i} \lambda }
\bigg|^2
2 \pi \delta( \omega_1' + \omega_2' )
	\frac{ {\rm d}^3\kappa_1\,{\rm d}^3\kappa_2 }{ 2 } 
%
%\nonumber\\
%&& \qquad \times 
	\label{eq:result-PA-1}
\end{eqnarray}
(the factor $1/2$ comes again from the Bose symmetry). This yields, to the fourth order in $d$,
the power
\begin{eqnarray}
P_2 
%&=& \int\!{\rm d}w_2 \, \hbar ( \omega_1 + \omega_2 )
%\label{eq:}\\
%&=&
=\frac{ d^4 }{ \pi^3 \hbar } 
\int &&\!{\rm d}^3\kappa_1\,{\rm d}^3\kappa_2\,
\, {\rm e}^{ - 2 ( k_1 + k_2 ) z }
\frac{  
|\vec{k}_1 \cdot \vec{k}_2|^2 }{ k_1 k_2 } 
\nonumber\\
&& {} \times
( \omega_1 + \omega_2 )
\mathop{\rm Im}R( \omega_1 )
\mathop{\rm Im}R( \omega_2 )
\frac{ \Omega^2 \, \delta( \omega_1' + \omega_2' )
}{ 
( \Omega + \omega_1' )^2
( \Omega + \omega_2' )^2
}~.
	\label{eq:recover-PA}
\end{eqnarray}
A simplified evaluation is reviewed 
in~\ref{a:power-Barton-order-v4}, leading to a scaling $\sim v^4$
for a small velocity [Eq.(5.4) of Ref.\cite{Barton2010b}]:
\begin{equation}
P_2 \simeq 
\frac{ 9 }{ 512 \pi }
\frac{ \hbar v^4 \alpha^2 \omega_p^4 \Gamma^2 }{ \omega_S^8 z^{10}}~.
\label{eq:Barton-PA-in-scaling-limit}
\end{equation}
% CH
Such a scaling with velocity was also found within fluctuation
electrodynamics~\cite{Dedkov2012,Pieplow2013,Intravaia2014},
although the numerical prefactor is different. 
Eqs.(\ref{eq:recover-PA}, \ref{eq:Barton-PA-in-scaling-limit})
coincide exactly with one term in Barton's results,
called there $P_{A}$ (see Eq.(5.4) of Ref.\cite{Barton2010b}). 
It is sub-leading, however, 
compared to another contribution (called $P_{B}$) that
scales as ${\cal O}( v^2 )$. Such a leading velocity dependence was
also put forward in Refs.\cite{Scheel2009,Milonni2013}. We analyze
the origin of the $P_B$ contribution of Ref.\cite{Barton2010b}
in Sec.\ref{s:generic-path} where the dependence on the atomic trajectory
is pointed out.
The calculations of 
Refs.\cite{Scheel2009,Intravaia2014} are reviewed
in Secs.\ref{s:macro-QED}, \ref{s:non-eq-FD}, respectively.

With respect to the scaling with the frequency parameters 
for the material response in Eq.(\ref{eq:Barton-PA-in-scaling-limit}), 
a similar behaviour
has been observed in previous work on the metal-vacuum surface 
where the 
plasmon resonance is at $\omega_S = \omega_p / \sqrt{2}$  \cite{Pendry1997,
ZuritaSanchez2004,Volokitin2007}. 
The combination $\Gamma / \omega_p^2$ is then proportional to the 
specific resistance of the metal. Only quasi-DC parameters are
relevant for these processes, the spectrum of the plasmon pairs
being confined to a region of width $\sim v / z$ around zero
frequency.

%%%%%%%%%

Since it will play an important role for a generic
trajectory [Sec.\ref{s:generic-path}], let us also discuss here
the one-photon power $P_{1}$.
From Eq.(\ref{eq:def-excitation-power}) we have that to the second order in $d$
it is connected with the 
squared amplitude
$|c_1^{(1)}( t )|^2$, leading to the differential excitation rate
\begin{equation}
{\rm d}w_1 = 
\sum_{\vec{\eta}}
\frac{ d^2 \, |\vec{\eta} \cdot \vec{k}|^2 |\phi_{\kappa}|^2 }{ 
\hbar^2 }
2 \pi \delta( \Omega + \omega' )
\, {\rm d}^3 \kappa~.
	\label{eq:diff-prob}
\end{equation}
Summing over all final photon states, we recover exactly the
excitation rate $\gamma_g$ obtained Eqs.(\ref{eq:0-level-shift-and-rate},
\ref{eq:Golden-Rule-lifetime}). An exponentially small scaling
with velocity still holds when the excitation energy is included
in the evaluation of
$P_1^{(2)} = \int\!{\rm d}w_1 \hbar( \Omega + \omega )$ (the superscript indicates again
the perturbative order). A consistent perturbative comparison with $P_{2}$ needs, however, 
a calculation up to the fourth order
in the coupling constant.
To evaluate the correction
$P_1^{(4)}$ to this power in the
next order, we consider the mixed term 
$2 \mathop{\rm Re} \, [c_1^{(1)*}( t )c_1^{(3)}( t )]$
and focus on its most divergent part,
namely the one increasing with $t$. We find a decrease of the 
emission rate:
\begin{equation}
P_1^{(4)} \approx - \gamma_g t P_1^{(2)}~.
\label{eq:}
\end{equation}
This suggests the
resummation $P_1 \approx P_1^{(2)} 
\, {\rm e}^{ - \gamma_g t }$, as expected by the instability
of the ground state. 
This shows that, also to the fourth order, the one-photon power is
exponentially suppressed
leaving, in the case of a constant velocity, $P_{2}$ and then a force $F\sim
v^{3}$ as the only relevant contribution to quantum friction.
In Sec.\ref{s:generic-path} we analyze how these results generalize for a more realistic case where the atom,
initially at rest, is accelerated to a constant velocity ${\bf v}$.

%%%%%%%%%%%%%%%%%%%%%%%%%%%%%%

\subsection{Average radiation force}
\label{s:ave-force}

Before proceeding, it is very instructive to directly evaluate the
frictional force given in Eq. (\ref{eq:force-operator}).
We consider here
the expectation value 
$\mathbf{F}(t) = \langle \Psi (t) | \hat{\mathbf{F}}(t) | \Psi (t)\rangle$
for an atom in uniform motion parallel to the surface. We shall use
again the expansion of the atom+field state $|\Psi (t) \rangle$
up to the third order of the interaction given in Eq.(\ref{eq:state-expansion-FD}). 
The nonzero matrix elements of the force 
operator can be derived from Eqs.(\ref{eq:ge-element-for-F}--%
\ref{eq:eg-element-for-F}): one just needs to replace the prefactor
${\rm i}$ in these equations by ${\bf k}$ or ${\bf k}_{1,2}$.
They yield the average force in the form
\begin{eqnarray}
\mathbf{F}(t) &=&
2 \mathop{\rm Re} 
\Big\{ 
\sum_{\vec{\eta}} \!\int\!{\rm d}^3\kappa\,
\langle g,\mathrm{vac}| \mathbf{\hat{F}}(t) | \vec{\eta},\kappa \rangle
% CH
( c_1^{(1)}( t ) + c_0^{(2)*}( t )c_1^{(1)}( t ) + c_1^{(3)}( t ) )
% CH
\nonumber\\
&& +
\frac12 
\sum_{\vec{\eta}} \!
\int\!{\rm d}^3\kappa \,{\rm d}^3\kappa_1 \,{\rm d}^3\kappa_2 \,
\langle \vec{\eta}, \kappa | \mathbf{\hat{F}}(t) 
	| g,\kappa_{1}\kappa_{2} \rangle 
c_1^{(1)*}( t )
c_2^{(2)}( t )
\Big\}~,
	\label{eq:expansion-ave-force-3}
\end{eqnarray}
where we included products of amplitudes up to order three.

After a straightforward calculation based on 
Eqs.(\ref{eq:c1-amplitude}, \ref{eq:c2-amplitude},
\ref{eq:c0-amplitude-2}, \ref{eq:c3-adiabatically}) for the
amplitudes $c_n^{(p)}( t )$ [details in \ref{a:ave-force}],
we find that the average force in the long-time limit 
$t \rightarrow \infty$ can be written as  $\mathbf{F} = \mathbf{F}^{( 2 )} + \mathbf{F}^{( 4 )}$.
The first term
\begin{eqnarray}
\mathbf{F}^{( 2 )}
&=&
- \frac{\hbar \alpha \Omega }{\pi }\!
\int\!\mathrm{d}^{3}\kappa \,
\mathbf{k}\,k\,\mathrm{e}^{-2kz}
\mathop{\rm Im}R( \omega )
\delta( \Omega + \omega' )  
\label{eq:correc-I-f}
\end{eqnarray}
is second order in the coupling constant and has a simple interpretation: 
it is the recoil due to the emission of a photon. 
This process is accompanied by the excitation of the atom 
(Cherenkov-Vavilov radiation) and happens
at the differential rate ${\rm d}w_1$ of Eq.(\ref{eq:diff-prob}). 
With every emission,
the atom receives a momentum $- \hbar {\bf k}$
opposite to the plasmon momentum.
The resulting force acting on the atom is
$
\mathbf{F} = - \int \!{\rm d}w_{1} \, \hbar \mathbf{k} 
%	\label{eq:force-from-recoil}
$
which coincides with Eq.(\ref{eq:correc-I-f}).

As explained in~\ref{a:ave-force}, the fourth-order contribution 
to the force can be presented in the form
\begin{eqnarray}
\mathbf{F}^{( 4 )} &=&
- \frac{ \mathbf{v} }{ v^{2} } P_{2} 
- \gamma_{g} t \,\mathbf{F}^{( 2 )} 
- \nabla_{\mathbf{v}}( \gamma_{g} \delta E_{g} )  
+ \mbox{other exp. small terms}
\label{eq:correc-III-f}
\end{eqnarray}
where $\gamma_{g}(v)$ and $\delta E_{g}(v)$ are the
relaxation rate and Lamb shift of the ground state,
respectively, Eq.(\ref{eq:0-level-shift-and-rate}).
Recall that $\gamma_{g}(v)$ arises from 
quantum Cherenkov-Vavilov radiation and is exponentially small.
This is also true for the second-order force $\mathbf{F}^{( 2 )}$ 
{}[Eq.(\ref{eq:correc-I-f})] because 
the resonance condition $\Omega + \omega' = 0$ 
involves the same threshold as 
Eq.(\ref{eq:gamma-g-exp-small}) for $\gamma_g$. 
The `other exp.\ small' terms not written
explicitly in Eq.(\ref{eq:correc-III-f}) have a similar origin. 

The important result is that the force~(\ref{eq:correc-III-f})
gives the leading order for small velocities and
involves the two-photon power $P_{2}$ obtained in Eq.(\ref{eq:recover-PA}).
This means that for a uniformly moving atom,
the average radiation force  
starts like ${\cal O}( v^3 )$, which is coherent with the 
radiated power obtained in the previous section.
The force calculation thus provides an
independent confirmation that two-photon rather than one-photon emission 
(plus atomic excitation) is the dominant loss process.
A comparison with the results of Dedkov and 
Kyasov~\cite{Dedkov2008} and Intravaia et al.~\cite{Intravaia2014}
is made in \ref{a:power-Barton-order-v4}: agreement up to a 
numerical factor is found when the atomic transition is off-detuned
with respect to the surface plasmon resonance, $\Omega \ll \omega_S$.
The dependence on distance involves the steep power law
$F \sim 1 / z^{10}$.

%%%%%%%%%%

\section{Accelerating the atom and subsequent radiation}
\label{s:generic-path}

In this section, we consider atomic trajectories that are accelerated
over a finite duration before reaching their final velocity
${\bf v}$. This material generalizes the calculation of 
Ref.\cite{Barton2010b} of the two-photon process where a second term
(called $P_B$) in the two-photon emission was found that
scales with ${\cal O}( v^2 )$ in velocity. The main result is
that the term $P_B$ depends sensitively on the way the atom is
accelerated. Ref.\cite{Barton2010b} is only recovered for a 
`sudden boost' (infinitely short duration), while in the opposite
or `adiabatic' limit, $P_B$ becomes strongly suppressed. 
The scaling with velocity ${\cal O}(v^2)$
is maintained, though. 

To interpret this behaviour, we have also evaluated the one-photon power $P_1$ and 
found, quite surprisingly, that for accelerated trajectories it is not exponentially small (as in the previous section),
it is negative, and exactly cancels the two-photon emission $P_B$. This
suggests the following picture: the acceleration stage creates
a finite occupation $p_e \sim v^2$ 
of the excited state (including one photon).
The excitation process is qualitatively similar
to the `acceleration-induced radiative excitation of ground-state 
atoms' analyzed by Barton and Calogeracos~\cite{Barton2008}.
Subsequently, this `real' rather than `virtual' excitation decays
resonantly into another photon. The resonance condition fixes the
energy of the second photon, so that the radiative power 
captured by the term $P_B \sim p_e \gamma_e \hbar \Omega$ 
is balanced by a decaying excitation
probability (negative $P_1$). 

The calculation proceeds by working out the probability amplitudes,
starting with the one-photon amplitude 
\begin{eqnarray}
c_1^{(1)}( t )
&=& 
- \frac{ d ( \vec{\eta} \cdot
\vec{ k }^*) \phi_{\kappa}^* }{ \hbar }
\int\limits_{-\infty}^{t}\!{\rm d}t_1 
\, {\rm e}^{ {\rm i}(\Omega + \omega) t_1 } 
\, {\rm e}^{ -{\rm i}{\bf k} \cdot {\bf r}(t_1) } ~,
\label{eq:def-excitation-amplitude}
\end{eqnarray}
where the matrix element~(\ref{eq:ge-element-for-F}) of the atom-field
coupling was used. Recall the compact notation $\kappa = {\bf k}\omega$
for the photonic modes and note that we have kept a generic atomic path 
${\bf r}( t_1 )$ under the integral. 
The $t_1$-integral appearing here will be denoted
$\mathcal{A}( e, \kappa; t)$ and discussed in detail in
Sec.\ref{s:first-order-generic-path}. We prove there that at large
times (once the launch is completed), the amplitude takes the form
\begin{equation}
t \gg \tau: \quad
c_1^{(1)}( t ) \approx
- \frac{ d ( \vec{\eta} \cdot
\vec{ k }^*) \phi_{\kappa}^* }{ \hbar }
\bigg\{
\frac{ {\rm e}^{ {\rm i} (\Omega + \omega') t }
}{
{\rm i} ( \Omega + \omega' ) }
+ 
\mathcal{B}_{e, \kappa }
\bigg\}~,
\label{eq:result-c1-Barton}
\end{equation}
where the first term is the same as for a constant-velocity
path [Eq.(\ref{eq:c1-amplitude})]. We interpret the second term
as a non-adiabatic excitation process whose amplitude is approximately
(for small velocity) given by
\begin{equation}
\mathcal{B}_{e, \kappa } \approx
\frac{ 
{\rm i} \,( {\bf k} \cdot {\bf v} )
\,{\rm e}^{ - {\rm i} {\bf k} \cdot {\bf r}( 0 ) } 
}{
(\Omega + \omega)^2 }
\Sigma( (\Omega + \omega) \tau )~.
	\label{eq:result-Be-Barton-0}
\end{equation}
The dimensionless factor $\Sigma( (\Omega + \omega) \tau )$ depends on
the specific shape of the path. It is proportional to the Fourier
transform of the acceleration [Eq.(\ref{eq:excite-as-Fourier})] 
and decays to zero when the product of
duration $\tau$ and frequencies $\Omega + \omega$ is much larger
than unity. In the opposite limit (`sudden acceleration'),
$\Sigma( (\Omega + \omega) \tau ) \to 1$.

In the next order of perturbation theory, we deal with the two-photon
amplitude
\begin{eqnarray}
c^{(2)}_2( t ) &=& 
- \frac{ {\rm i} }{ \hbar }
\sum_{\vec{\eta}}\int\!{\rm d}\kappa
\int\limits_{-\infty}^{t}\!{\rm d}t_2\,
c_1^{(1)}( t_2 )
\langle g; \kappa_1, \kappa_2 | 
\hat{V}(t_2) 
| \vec{\eta};  \kappa \rangle
\label{eq:amplitude-c2}
\\
&=&
\frac{ d^2 }{ \hbar^2 }
( \vec{ k }_1 \cdot \vec{ k }_2 )^*
\phi_{\kappa_1}^* \phi_{\kappa_2}^*
\int\limits_{-\infty}^{t}\!{\rm d}t_2\,
\mathcal{A}( e, \kappa_1; t_2 ) 
\, {\rm e}^{ {\rm i}(- \Omega + \omega_2) t_2 } 
\, {\rm e}^{ -{\rm i}{\bf k}_2 \cdot {\bf r}(t_2) } 
+ \{ 1 \leftrightarrow 2 \}~.
\nonumber
\end{eqnarray}
The additional
term denoted $\{ 1 \leftrightarrow 2 \}$ makes this expression symmetric
under plasmon exchange. 
The $t_2$-integral written here 
will be called $\mathcal{M}_{12}( t )$
similar to Ref.\cite{Barton2010b}.
We find (Sec.\ref{s:second-photon}) for this two-photon amplitude 
the asymptotic form
\begin{equation}
t \gg \tau: \quad
{\cal M}_{12}( t ) \approx
-
\frac{ 
{\rm e}^{ {\rm i} ( \omega_1' + \omega_2' ) t }
 }{ ( \Omega + \omega_1') (\omega_1' + \omega_2') }
+
\frac{ \mathcal{B}_{e, \kappa_1 } }{ {\rm i}
( \omega_2' - \Omega ) }
{\rm e}^{ {\rm i}(\omega_2' - \Omega) t } 
+ {\rm cst.}~,
\label{eq:result-c2-Barton}
\end{equation}
where the ${\rm cst.}$ denotes $t$-independent terms. The first term
again recovers the previous constant-velocity result
from Eq.(\ref{eq:c2-amplitude}).
The second term is proportional to the non-adiabatic
excitation amplitude 
${\cal B}_{e,\kappa}$ [Eqs.(\ref{eq:result-c1-Barton},
\ref{eq:result-Be-Barton-0})] that appeared in the first order.

The power of two-photon emission from Eq.(\ref{eq:def-two-photon-power}) 
is proportional to 
$|\mathcal{M}_{12}( t ) + \mathcal{M}_{21}( t )|^2 / t$
in the large-$t$ limit. This is calculated in Eq.(\ref{eq:result-M})
below. In Sec.\ref{s:tentative-interpretation},
we finally discuss the rate of change of the energy stored 
in the excited state $| \vec{\eta}, \kappa \rangle$ 
(the one-photon power
$P_1$ defined in Eq.(\ref{eq:def-excitation-power})) and show that it
balances exactly the ${\cal O}( v^2 )$ contribution to the two-photon
power. 

%%%%%%%%%%%%%

\subsection{Exciting the atom: the one-photon amplitude}
\label{s:first-order-generic-path}

\begin{figure}[bhtp]
\centering
\includegraphics[height=48mm]{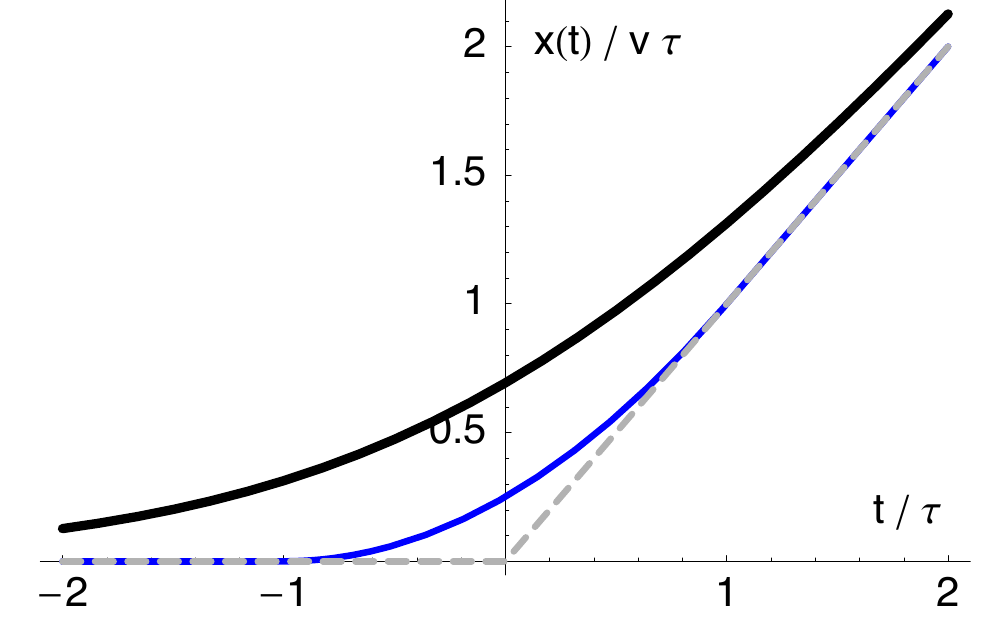} % requires the graphicx package
\caption[]{
(Color online) Three possible atomic trajectories: smooth continuous boost (thick black line), linear acceleration ramp (thin blue line), and instantaneous boost (dashed gray line). See also the table below for more detail.
   \\[1em]
	\begin{tabular}{c|c|c}
	     Curve             & $\dot x( t )$    &   $\ddot x( t )$ \\
	\hline
	thick black & $v / (1 + {\rm e}^{ -t / \tau })$      &  $v \tau^{-1} / (2 + 2 \cosh t / \tau )$ \\
	thin blue   &   $\left\{ \begin{array}{ll} 0 & \mbox{for $t < -\tau$} \\ 	( t + \tau ) v / (2\tau)  & \mbox{for $-\tau < t < \tau$} \\  v & \mbox{for $t>\tau$} \end{array} \right.$ & 
	$\left\{ \begin{array}{ll} 0 & \mbox{for $t < -\tau$} \\  v / (2\tau)  & \mbox{for $-\tau < t < \tau$} \\  0 & \mbox{for $t>\tau$} \end{array} \right.$ \\
	dashed gray & $\left\{ \begin{array}{ll} 0 & \mbox{for $t<0$} \\ v & \mbox{for $t>0$} \end{array} \right.$ 	 & $v \delta ( t )$ \\
	\end{tabular}	
}
   \label{fig:paths}
\end{figure}

Let us consider the first step of the physical process described above. The one-photon
transition amplitude is proportional to 
\begin{equation}
\mathcal{A}( e, \kappa; t) 
= \int\limits_{-\infty}^{t} \!{\rm d}t_1 \; {\rm e}^{i(\Omega+\omega) t_1} 
{\rm e}^{-i {\bf k} \cdot {\bf r}(t_1)}.
\label{def:amplitude-c1-1}
\end{equation}
A few general properties of the first-order amplitude can be secured without specifying a particular path.
We only require that ${\bf r}( t )$ changes its velocity around 
$t = 0$ with a typical duration $\tau$. We also 
assume that the origin of the coordinate system is chosen such that 
for $t \gg \tau$, we have ${\bf r}( t ) \approx {\bf v} t$ 
(see Fig.\ref{fig:paths} for a sketch).
Let us focus first 
on $t > 0$. We split the integral into $-\infty < t_1 \le 0$ and
$0 \le t_1 \le t$, leading to a natural decomposition
\(
\mathcal{A}( t ) = \mathcal{A}_- + \mathcal{A}_+( t )
\).
For the term $\mathcal{A}_+( t )$, we perform a partial integration 
after subtracting and adding ${\bf v} t_1$ in the exponent 
[see Eq.(\ref{eq:def-excitation-amplitude})]. This leads to
\begin{equation}
\mathcal{A}_+( t ) = 
\left.
\frac{ {\rm e}^{ {\rm i} (\Omega + \omega) t_1 }
\,{\rm e}^{ - {\rm i} {\bf k} \cdot {\bf r}( t_1 ) } 
}{
{\rm i} ( \Omega + \omega - {\bf k} \cdot {\bf v} ) }
\right|^{t}_{0}
+
\int\limits_{0}^{t}\!{\rm d}t_1
\frac{ {\bf k} \cdot ( \dot {\bf r} - {\bf v} ) }{
\Omega + \omega - {\bf k} \cdot {\bf v} }
{\rm e}^{ {\rm i} (\Omega + \omega) t_1 }
\,{\rm e}^{ - {\rm i} {\bf k} \cdot {\bf r}( t_1 ) } ~.
\label{eq:p-Int-result-for-A}
\end{equation}
The advantage of this representation is that the first term yields
what we call the `adiabatic limit'%
\footnote{%
This is not a fully adiabatic expression since the denominator
contains ${\bf v}$ instead of the instantaneous atomic velocity
$\dot{\bf r}( t )$.%
}
\begin{equation}
\mathcal{A}_+^{({\rm ad})}( e, \kappa; t ) = 
\frac{ {\rm e}^{ {\rm i} (\Omega + \omega) t }
\,{\rm e}^{ - {\rm i} {\bf k} \cdot {\bf r}( t ) } 
}{
{\rm i} ( \Omega + \omega - {\bf k} \cdot {\bf v} ) }
\label{eq:def-adiabatic-limit}
\end{equation}
which is nothing but the term we found previously for a trajectory
with constant velocity [see Eq.(\ref{eq:c1-amplitude})].
In the integrand in Eq.(\ref{eq:p-Int-result-for-A}), 
the difference $\dot {\bf r} - {\bf v}$ vanishes
as soon as the atom has reached a constant velocity. 
Hence the
integral approaches a constant for $t \gg \tau$. In this limit, we
therefore proved that the excitation amplitude takes the form
\begin{equation}
t \gg \tau: \qquad
\mathcal{A}( e, \kappa; t ) =
\mathcal{A}_+^{({\rm ad})}( e, \kappa; t ) +
\mathcal{B}_{e, \kappa }
+ o( 1 )~,
\label{eq:asymptotic-split-A}
\end{equation}
where $\mathcal{B}_{e, \kappa }$ can be read off by adding
$\mathcal{A}_-$ to the remaining terms in Eq.(\ref{eq:p-Int-result-for-A}).
The error $o( 1 )$ in Eq.(\ref{eq:asymptotic-split-A}) is made of terms that 
vanish in the limit $t \gg \tau$. 
A similar manipulation can be applied when $t \le 0$.
No splitting and subtraction are needed, and we get
\begin{equation}
t \le 0: \qquad
\mathcal{A}( e, \kappa; t ) = 
\frac{ {\rm e}^{ {\rm i} (\Omega + \omega) t }
\,{\rm e}^{ - {\rm i} {\bf k} \cdot {\bf r}( t ) } 
}{
{\rm i} ( \Omega + \omega ) }
+
\int\limits_{-\infty}^{t}\!{\rm d}t_1
\frac{ {\bf k} \cdot \dot {\bf r} }{
\Omega + \omega }
{\rm e}^{ {\rm i} (\Omega + \omega) t_1 }
\,{\rm e}^{ - {\rm i} {\bf k} \cdot {\bf r}( t_1 ) } ~.
\label{eq:p-Int-result-for-A-early}
\end{equation}
The first term may again be called `adiabatic' 
and does not contain any contribution 
from the lower limit
$t = - \infty$ because we assume that the atom-field coupling is switched
off there. Note also that there is no Doppler shift in the frequency
denominator.
The second term vanishes for $t \ll - \tau$ when the atom
is still at rest.

%%%%%%%%

\subsection{Emitting the second photon}
\label{s:second-photon}

Here, we focus on the two-photon amplitude in the long-time
limit $t \to \infty$. Using the time scale $\tau$ for the `acceleration
stage' of the atomic trajectory, we assume more specifically 
$t \gg \tau$. Let us introduce a time $t_a$ with the property
$\tau \ll t_a \ll t$ such that the atomic velocity is $\dot {\bf r}( t_a ) 
= {\bf v}$. We split the integration range of the $t_2$-integral
in Eq.(\ref{eq:amplitude-c2}) into $- \infty < t_2 < t_a$ 
and into $t_a < t_2 < t$ obtaining $\mathcal{M}_{12}( t ) = 
\mathcal{M}_- + \mathcal{M}_+( t )$ (and similarly for $\mathcal{M}_{21}( t ) $).
We shall see that only the probability $|\mathcal{M}_+( t )|^2$ contains
terms growing linearly with $t$ and then contributing to the radiated power, while the rest tends towards a constant.
The contribution ${\cal M}_+( t )$ can be evaluated by inserting the
asymptotic form~(\ref{eq:asymptotic-split-A}) for the amplitude
${\cal A}( e, \kappa_1; t )$ into the integral:
\begin{eqnarray}
{\cal M}_+( t ) 
&=& \int\limits_{t_a}^{t}\!{\rm d}t_2\,
\left(
\mathcal{A}_+^{({\rm ad})}( e, \kappa_1; t_2 ) +
\mathcal{B}_{e, \kappa_1 }
\right)
{\rm e}^{ {\rm i}(\omega_2 - \Omega ) t_2 } 
\, {\rm e}^{ -{\rm i}{\bf k}_2 \cdot {\bf r}(t_2) } 
\nonumber
\\
&=& 
{\cal M}_+^{(A)}( t ) + {\cal M}_+^{(B)}( t ).
\label{eq:split-Mplus-in-A-B}
\end{eqnarray}
Because, for $t > t_a$, the atomic velocity is constant and the first term 
evaluates to 
\begin{equation}
{\cal M}_+^{(A)}( t ) = 
-
\left.
\frac{ 
{\rm e}^{ {\rm i} ( \omega_1' + \omega_2' ) t_2 }
 }{ ( \Omega + \omega_1') (\omega_1' + \omega_2') }
\right|^{t}_{t_a},\quad (\omega_i' = \omega_i - {\bf k}_1 \cdot {\bf v})
\label{eq:}
\end{equation}
we recover the result for a constant velocity [cf. Eq.(\ref{eq:c2-amplitude})].
We symmetrize under photon exchange, square and get in the
limit $t \gg t_a$:
\begin{equation}
\left| {\cal M}_+^{(A)}( t ) + \{ 1 \leftrightarrow 2 \} \right|^2
= 2\pi ( t - t_a ) 
\delta( \omega_1' + \omega_2' )
\frac{ (2\Omega + \omega_1' + \omega_2')^2 }{ 
(\Omega + \omega_1')^2 (\Omega + \omega_2')^2 }
\label{eq:square-leading-to-PA}~,
\end{equation}
where the sum $\omega_1' + \omega_2'$ in the numerator can of course
be omitted. 
The second term ${\cal M}_+^{(B)}( t )$ in Eq.(\ref{eq:split-Mplus-in-A-B})
is an elementary integral as well:
\begin{equation}
{\cal M}_+^{(B)}( t ) = 
\frac{ \mathcal{B}_{e, \kappa_1 } }{ {\rm i}
( \omega_2' - \Omega ) }
\left.
{\rm e}^{ {\rm i}(\omega_2' - \Omega) t_2 } 
\right|^{t}_{t_a}~.
\label{eq:result-M-B}
\end{equation}
We symmetrize again and identify the individual squares as the leading terms:
\begin{equation}
\left| {\cal M}_+^{(B)}( t ) +
	\{ 1 \leftrightarrow 2 \} \right|^2 =
2\pi ( t - t_a ) 
\delta( \omega_2 - \Omega )
|\mathcal{B}_{e, \kappa_1 }|^2 
+
\{ 1 \leftrightarrow 2 \}~.
\label{eq:square-leading-to-PB}
\end{equation}
It can be checked that the `mixed terms' in the squared amplitude
lead to contributions that either oscillate or
tend to constants as $t \to \infty$
(no inverse squares like $1/(\omega_1'+\omega_2')^2$ appear).
This also holds true for the mixed terms involving one factor
${\cal M}_-$
(see \ref{a:piecewise} for more details).
Finally we get
\begin{eqnarray}
\lim_{t \rightarrow \infty} \frac{|{\cal M}_{12}(t) + {\cal M}_{21}(t)|^2}{t} 
&=& 2 \pi \delta(\omega_1' + \omega_2')  
\frac{ 4 \Omega^2 }{ (\Omega + \omega_1' )^2 (\Omega + \omega_1' )^2}
\nonumber
\\
&& 
+ 
2 \pi \delta(\omega_2' - \Omega) 
\left| \mathcal{B}_{e,\kappa_1 } \right|^2 
+ \{ 1 \leftrightarrow 2\}~.
\label{eq:result-M}
\end{eqnarray}

%%%%%%%%%%%%%%%

From the above calculation, one can see that the first line 
of Eq.(\ref{eq:result-M}) involves only the constant-velocity
part of the atomic path and hence does not depend on the way
the atom is put into motion. As was reviewed in 
Sec.\ref{s:Barton-PA-1}
and~\ref{a:power-Barton-order-v4},
this term contributes the amount $P_A \sim v^4$ 
to the two-photon emission.
The second line of Eq.(\ref{eq:result-M}) leads to 
\begin{equation}
P_B = 
\frac{\alpha^2 \Omega^2 }{ 4 \hbar }
\int\!{\rm d}^3\kappa_1 \, {\rm d}^3\kappa_2 \, 
(\omega_1 + \omega_2)
| \vec{ k }_1 \cdot \vec{ k }_2 |^2
|\phi_1 \phi_2|^2
2\pi 
\delta( \omega_2 - \Omega )
|\mathcal{B}_{e, \kappa_1 }|^2
\label{eq:power-PB-2}
\end{equation}
and scales with $v^2$ for small velocity
{}[\ref{a:power-Barton-order-v2}].
This arises because
the amplitude $\mathcal{B}_{e, \kappa } \sim v$, as we now
show.

We start with the general expression collected from the terms in 
Eqs.(\ref{eq:p-Int-result-for-A}, \ref{eq:p-Int-result-for-A-early})
that become constants for large $t \gg \tau$:
\begin{eqnarray}
\mathcal{B}_{e, \kappa } &=&
\frac{ {\rm e}^{ - {\rm i} {\bf k}\cdot {\bf r}( 0 ) } }{ 
{\rm i} ( \Omega + \omega) }
+
\int\limits_{-\infty}^{0}\!{\rm d}t_1
\frac{ {\bf k} \cdot \dot {\bf r} }{
\Omega + \omega }
{\rm e}^{ {\rm i} (\Omega + \omega) t_1 }
\,{\rm e}^{ - {\rm i} {\bf k} \cdot {\bf r}( t_1 ) } 
\nonumber
\\
&& {}
-
\frac{ {\rm e}^{ - {\rm i} {\bf k}\cdot {\bf r}( 0 ) } }{ 
{\rm i} ( \Omega + \omega - {\bf k} \cdot {\bf v}) }
+
\int\limits_{0}^{\infty}\!{\rm d}t_1
\frac{ {\bf k} \cdot ( \dot {\bf r} - {\bf v} ) }{
\Omega + \omega - {\bf k} \cdot {\bf v} }
{\rm e}^{ {\rm i} (\Omega + \omega) t_1 }
\,{\rm e}^{ - {\rm i} {\bf k} \cdot {\bf r}( t_1 ) } ~.
\label{eq:general-form-for-Ae}
\end{eqnarray}
The first terms in the two lines sum to
\begin{equation}
\frac{ {\rm e}^{ - {\rm i} {\bf k}\cdot {\bf r}( 0 ) } }{ 
{\rm i} ( \Omega + \omega) }
-
\frac{ {\rm e}^{ - {\rm i} {\bf k}\cdot {\bf r}( 0 ) } }{ 
{\rm i} ( \Omega + \omega - {\bf k} \cdot {\bf v}) }
=
\frac{ ( - {\bf k} \cdot {\bf v} )\,
{\rm e}^{ - {\rm i} {\bf k}\cdot {\bf r}( 0 ) } 
	}{ 
{\rm i} ( \Omega + \omega) ( \Omega + \omega - {\bf k} \cdot {\bf v}) }~.
\label{eq:Barton-amplitude}
\end{equation}
This is also what is found for Ref.\cite{Barton2010b}'s `kink trajectory' 
where 
$\dot {\bf r} = 0$ for $t < 0$ and $\dot {\bf r} = {\bf v}$ for $t > 0$.
The remaining integrals are bounded by ${\cal O}( k v \tau 
/ ( \Omega + \omega ) )$ so that their contribution to the amplitude 
$\mathcal{B}_{e, \kappa }$ is also linear in the velocity
$v$. 
To proceed, we apply another partial integration to the two integrals
in Eq.(\ref{eq:general-form-for-Ae}). Summing the results
and expanding for small $v$, one gets

\begin{eqnarray}
&&
\int\limits_{-\infty}^{0}\!{\rm d}t_1
\frac{ {\bf k} \cdot \dot {\bf r} }{
\Omega + \omega }
{\rm e}^{ {\rm i} (\Omega + \omega) t_1 }
\,{\rm e}^{ - {\rm i} {\bf k} \cdot {\bf r}( t_1 ) } 
+
\int\limits_{0}^{\infty}\!{\rm d}t_1
\frac{ {\bf k} \cdot ( \dot {\bf r} - {\bf v} ) }{
%\Omega + \omega' }
\Omega + \omega - {\bf k} \cdot {\bf v} }
{\rm e}^{ {\rm i} (\Omega + \omega) t_1 }
\,{\rm e}^{ - {\rm i} {\bf k} \cdot {\bf r}( t_1 ) } 
\nonumber
\\
&& \approx
\frac{ {\bf k} \cdot  {\bf v} 
\,{\rm e}^{ - {\rm i} {\bf k} \cdot {\bf r}( 0 ) } 
}{
{\rm i} (\Omega + \omega)^2 }
-
\frac{ 
{\rm e}^{ - {\rm i} {\bf k} \cdot {\bf r}( 0 ) } 
}{
{\rm i} (\Omega + \omega)^2 }
\!
\int\limits_{-\infty}^{\infty}\!{\rm d}t_1
{\bf k} \cdot \ddot {\bf r}
\,{\rm e}^{ {\rm i} (\Omega + \omega) t_1 }
\end{eqnarray}
where the first term cancels with~(\ref{eq:Barton-amplitude}) to
leading order in $v$. We are thus
left with the Fourier integral of the atomic acceleration
\begin{equation}
\mathcal{B}_{e, \kappa } \approx
-
\frac{ 
{\rm e}^{ - {\rm i} {\bf k} \cdot {\bf r}( 0 ) } 
}{
{\rm i} (\Omega + \omega)^2 }
\!
\int\limits_{-\infty}^{\infty}\!{\rm d}t_1
{\bf k} \cdot \ddot {\bf r}(t_{1})
\,{\rm e}^{ {\rm i} (\Omega + \omega) t_1 }=\frac{ 
{\rm i} \,( {\bf k} \cdot {\bf v} )
\,{\rm e}^{ - {\rm i} {\bf k} \cdot {\bf r}( 0 ) } 
}{
(\Omega + \omega)^2 }
\Sigma( (\Omega + \omega) \tau )~.
	\label{eq:excite-as-Fourier}
\end{equation}
which is the result announced in Eq.(\ref{eq:result-Be-Barton-0}). This already permits us to draw a conclusion for a generic trajectory
whose velocity is monotonously raised to its final value ${\bf v}$.
The Fourier transform of the acceleration then exists and is maximal
for $\Omega + \omega = 0$. If the acceleration occurs over a finite
duration ${\cal O}( \tau )$, the Fourier transform drops to zero when
the frequencies satisfy $(\Omega + \omega) \tau \gg 1$. On
physical grounds, it seems quite plausible that the acceleration 
stage takes more than a few femtoseconds while $\Omega$ is typically
in the visible range (and $\omega > 0$). This inequality is 
therefore amply satisfied, and the corresponding two-photon emission is
strongly suppressed. 

%%%%%%%%%%%%%%%%%%%%%%%%%%

It is interesting to examine different atomic trajectories ${\bf r}(t)$
and provide a quantitative analysis 
how the `acceleration stage' affects the final result for the two-photon
emission $P_B$. 
A trivial example to begin with is an inertial path with constant 
velocity. The acceleration is zero at all times, and 
Eq.(\ref{eq:excite-as-Fourier}) gives ${\cal B}_{e,\kappa} = 0$.
Hence the two-photon power $P_B = 0$, and only the ${\cal O}( v^4 )$
contribution called $P_A$ from Sec.\ref{s:Barton-PA-1} remains.

Our second example is Ref.\cite{Barton2010b}'s instantaneous boost, 
$\ddot {\bf r}(t) = {\bf v} \delta( t )$. The trajectory is plotted
as a dashed line in Fig.\ref{fig:paths}.
The Fourier integral gives 
\begin{equation}
|\mathcal{B}_{e, \kappa }|^2 \approx
\frac{ ({\bf k} \cdot {\bf v} )^2 }{ ( \Omega + \omega)^4 }.
\label{eq:Barton-probability-e}
\end{equation}
This result can also been inferred from 
Eq.(4.10) of Ref.\cite{Barton2010b} by writing it in the form
given in Eq.(\ref{eq:result-M}) (and taking into account the erratum).

As a third example, consider a linear velocity ramp, as illustrated by the
middle path in Fig.\ref{fig:paths}. Velocity and acceleration
are given in the caption: specifically the acceleration is constant
during an interval of length $2\tau$ (details of the full calculation
for this path can be found in~\ref{a:piecewise}).
The sinc-function resulting
from the Fourier integral~(\ref{eq:excite-as-Fourier}) then gives 
\begin{equation}
|\mathcal{B}_{e, \kappa }|^2 
\approx
\frac{ ({\bf k} \cdot {\bf v})^2 }{ ( \Omega + \omega)^4 }
\frac{ \sin^2[ ( \Omega + \omega )\tau ] }{ ( \Omega + \omega)^2\tau^2  }~.
\label{eq:piecewise-path}
\end{equation}
The second fraction reproduces Ref.\cite{Barton2010b}'s path in the limit $\tau \to 0$,
but gives a strong reduction in the opposite case. Since the typical
frequencies contributing to the 
integral~(\ref{eq:power-PB-2}) are
$\omega_1 \simeq \omega_S$ due to the plasmon pole,
$|\phi_1|^2 \sim \mathop{\rm Im} R( \omega_1 )$,
the power $P_B$ gets reduced
by a factor $1 / (\tau (\Omega + \omega_S))^2 \ll 1$. 
This result for the linear velocity ramp has been reproduced 
through a differently routed calculation by G. Barton (private communication).

Finally, let us consider the `smooth boost' plotted as a thick black
line in Fig.\ref{fig:paths}. The acceleration is given by an
infinitely differentiable function (see figure caption) whose maximum
value is $v / (4 \tau)$ and whose width is ${\cal O}( \tau )$.
Evaluating its Fourier transform, we get the probability
\begin{equation}
|{\cal B}_{e,\kappa}|^2 =
\frac{ ({\bf k} \cdot {\bf v})^2 }{ (\Omega + \omega)^4 }
\left[
\frac{\pi (\Omega + \omega) \tau }{\sinh[\pi (\Omega + \omega) \tau]}
\right]^2~.
\label{eq:accel}
\end{equation}
The second fraction in this expression shows that compared to the
`kink path', the power $P_B$ becomes exponentially small in the limit of 
an adiabatic boost $\tau \Omega \gg 1$. 

%%%%%%%%%%%%%%

%
Let us attach a physical meaning to the quantities calculated in the
previous sections.
Going back to the Schr\"odinger picture, the first-order amplitude
$c_1^{(1)}( t )$ [Eq.(\ref{eq:result-c1-Barton})] takes the
form
\begin{equation}
c_1^{(1)}( t ) \approx
- \frac{ d ( \vec{\eta} \cdot
\vec{ k }^*) \phi_{\kappa}^* }{ \hbar }
\bigg\{
\frac{ 1 }{
{\rm i} ( \Omega + \omega' ) }
+ 
\mathcal{B}_{e, \kappa }
\,{\rm e}^{ -{\rm i} (\Omega + \omega') t }
\bigg\}~.
%\label{eq:result-c1-Barton}
\end{equation}
The first term, independent of $t$, can be understood as being
part of the (`dressed') 
ground state $| g, {\rm vac} \rangle$ 
(still at zero energy), 
where the atom is surrounded by a (`virtual') cloud of photons 
(plasmons). The second term oscillates at the (bare) energy of
the 
excited state $| \vec{\eta}, \kappa \rangle$, but including
the Doppler shift ($\omega'$ instead of $\omega$). It can be
shown that the Hamiltonian of our atom+field theory can be
transformed to a time-independent form by going into a frame
moving with the atom. (Details are postponed to another paper.)
In this picture, the state $| \vec{\eta}, \kappa \rangle$
evolves freely at the frequency $\Omega + \omega'$. We therefore
conjecture that along a path with a time-dependent acceleration,
the amplitude $\mathcal{B}_{e, \kappa }$ describes
the `real' excitation of the atom+field system~\cite{Barton2008}. 
The required 
energy transfer is in heuristic agreement with the frequency
uncertainty arising from the finite duration of
the acceleration, as expressed in the Fourier 
integral~(\ref{eq:excite-as-Fourier}).
%
%%%%%%%%%%

We can also  define an 
excitation \emph{probability} (not a rate), 
summing over the plasmon states and
the three sublevels $| \vec{\eta} \rangle$
\begin{eqnarray}
p_e &=& \sum_{\vec{\eta}} 
\frac{ d^2 }{ \hbar^2 }
\int\!{\rm d}^3\kappa\,
|\vec{\eta} \cdot \vec{ k }|^2 |\phi_{\kappa}|^2 
| \mathcal{B}_{e, \kappa } |^2
\label{eq:}\nonumber\\
&=&
\frac{ \alpha \Omega }{ 2 \pi^2 }
\int\!{\rm d}^3\kappa\,
k \,{\rm e}^{ - 2 k z }
\mathop{\rm Im}R ( \omega )
\frac{ ({\bf k} \cdot {\bf v} )^2 }{ ( \Omega + \omega)^4 }
|\Sigma( (\Omega + \omega) \tau )|^2~.
\end{eqnarray}
where the $k$-integral can be performed, yielding $3 \pi v^2 / (4 z^5)$.

%%%%%%%%%%%

\subsection{Excitation power: a subtle cancellation to the fourth order}
\label{s:tentative-interpretation}

The previous analysis examined in detail all the components of the physical
process describing the acceleration, the excitation and the subsequent radiation of an atom
in motion near a surface. This was necessary in order to discern and generalize, to the case
of a generic trajectory, each single contribution to quantum friction.
In Ref.\cite{Barton2010b} the friction force ${\bf F}$
is calculated based on the identification with the two-photon
power loss, $P_2 = - {\bf v} \cdot {\bf F}$. 
For a generic trajectory this calculation led to $P_2=P_{A}+P_{B}$
where $P_{B}\sim {\cal O}(v^{2})$ gives the leading order.
Barton thus concludes that ${\bf F}\sim v$ for
small velocity~\cite{Barton2010b}, at least for the trajectory called 
instantaneous boost above.
In the last section we showed, however, that the value of $P_{B}$ depends on the detail of the trajectory: a smooth boost gives a significant reduction,
and a constant velocity simply leads to $P_{B} = 0$ (see also 
Sec.\ref{s:Barton-PA-1}).

Although physically sounding, the calculation based on $P_2$ is 
incomplete since it does not take into account
the power needed to create the excited state 
$| \vec{\eta}, \kappa \rangle$, called the one-photon power $P_1$
in Eq.(\ref{eq:def-excitation-power}). 
A similar omission in earlier work was criticized by Volokitin and Persson, 
see Ref.\cite{Volokitin2002}.
(For an analysis of the `internal energy'
appearing in $P_1$, see also Ref.\cite{Dedkov2008}, for example.)
The sum $P_1 + P_2$ translates the change in the total energy
of the evolving state $| \Psi( t ) \rangle$. This energy is not conserved,
since the interaction is time-dependent. For the same reason, only
the free Hamiltonian $\hat{H}_0 = \hat{H}_A + \hat{H}_F$ (atom and field)
is used to define the energy of the state
$| \Psi( t ) \rangle$:
\begin{equation}
\langle \Psi( t ) | \hat{H}_0 | \Psi( t ) \rangle \approx 
t (P_1 + P_2) + \ldots
\label{eq:def-energy-change}
\end{equation}
The power $P_1$ is calculated again
by pushing to the third order the one-photon amplitude
$\langle \vec{\eta}, \kappa | \Psi( t ) \rangle$,
called $c_1^{(3)}( t )$ 
in Eq.(\ref{eq:state-expansion-FD}).
This extension is necessary because at first order,  even with an acceleration phase, the excitation
rate is exponentially small [see Eq.(\ref{eq:gamma-g-exp-small})].
We focus again on the state 
sequence
$| g, {\rm vac} \rangle 
\to | \vec{\eta}, \kappa \rangle 
\to | g, \kappa_1 \kappa_2 \rangle 
\to | \vec{\eta}, \kappa \rangle$
passing via the two-photon state. 
(The sequence via the ground state $| g, {\rm vac} \rangle$
gives again exponentially small contributions.)

The calculation proceeds along lines similar to Sec.\ref{s:Barton-PA-1}. Perturbation theory yields the 
integral~(\ref{eq:c3-perturb-iteration}) where we insert now the expression for
$\langle g, \kappa_1 \kappa_2 | \Psi( t ) \rangle$ generalized to the case of a generic
trajectory
(proportional to the amplitude ${\cal M}( t )$ from
Sec.\ref{s:second-photon}).  
In addition to the constant velocity result we obtain 
a correction $\delta c_1^{(3)}( t )$ to the amplitude
coming from the second
term in Eq.(\ref{eq:result-c2-Barton}) and its symmetrized partner.
We consider that interaction times in the interval $\tau < t_3 < t$ 
give the main contribution and approximate
\begin{eqnarray}
\delta c_1^{(3)}( t ) &\approx&
\frac{ d^3 }{ \hbar^3 }
\int\!{\rm d}^3\kappa_1 \!
( \vec{ k }_1 \cdot \vec{ k } )^*
|\phi_{\kappa_1}|^2 \phi_{\kappa}^*
(\vec{\eta} \cdot \vec{k}_1) 
\nonumber\\
&& \quad {} \times
\int\limits_{\tau}^{t}\!{\rm d}t_3
\bigg\{
\frac{ {\cal B}_{e, \kappa_1} 
\, {\rm e}^{ {\rm i} ( \omega' - \omega_1' ) t_3 }
}{ {\rm i} ( \omega' - \Omega - {\rm i} \lambda ) }
+
\frac{ {\cal B}_{e, \kappa} 
}{ {\rm i} ( \omega'_1 - \Omega - {\rm i} \lambda ) }
\bigg\}~.
\label{eq:}
\end{eqnarray}
The second term in the curly brackets gives rise to a linear increase
in time. 
The fourth-order approximation to the excited-state probability $| \langle \vec{\eta}, \kappa | \Psi( t ) \rangle |^2
\approx |c_1^{(1)}( t )|^2 + 2 \mathop{\rm Re}[ c_1^{(1)*}( t ) \delta c_1^{(3)}( t ) ]+ \ldots$
thus provides us with an excitation rate
\begin{eqnarray}
&&
\lim_{t\to\infty}
\frac{ | \langle \vec{\eta}, \kappa | \Psi( t ) \rangle |^2 
- |c_1^{(1)}( t )|^2 }{ t } 
\nonumber\\
&&
\approx
- \frac{ 2 d^4 }{ \hbar^4 }
\mathop{\rm Re}\bigg[
( \vec{\eta} \cdot \vec{ k }) 
|\phi_{\kappa}|^2
\int\!{\rm d}^3\kappa_1 \!
( \vec{ k }_1 \cdot \vec{ k } )^*
|\phi_{\kappa_1}|^2 
(\vec{\eta} \cdot \vec{k}_1)
\frac{ | {\cal B}_{e, \kappa} |^2
}{ {\rm i} ( \omega'_1 - \Omega - {\rm i} \lambda ) }
\bigg]~.
	\label{eq:result-excitation-rate}
\end{eqnarray}
Note that for this rate of change, we only need
the non-adiabatic amplitude
${\cal B}_{e, \kappa}$ in
the first-order expression $c_1^{(1)*}( t )$
{}[Eq.(\ref{eq:result-c1-Barton})], since
the other combinations give rise to oscillating contributions. 
The one-photon power~(\ref{eq:def-excitation-power}) becomes
\begin{equation}
P_1 = 
- \frac{ \alpha^2 \Omega^2 }{ 4\hbar }
\int\!{\rm d}^3\kappa\, {\rm d}^3\kappa_1 \!
( \Omega + \omega )
|\phi_{\kappa}|^2
|\phi_{\kappa_1}|^2 
| \vec{ k }_1 \cdot \vec{ k } |^2
| {\cal B}_{e, \kappa} |^2
2 \pi \delta( \omega'_1 - \Omega )~.
	\label{eq:result-one-photon-power}
\end{equation}
To the lowest order in velocity (recall that ${\cal B}_{e,\kappa}$
is proportional to $v$), we may drop the Doppler shift in
$\delta( \omega'_1 - \Omega )$. Now, it is possible to check that 
the one-photon power~(\ref{eq:result-one-photon-power}) 
exactly balances the contribution $P_B$ to the two-photon
power [Eq.(\ref{eq:power-PB-2})]. It leaves $P_{A}$ as the
only relevant contribution for the total dissipated power, even for
a particle path including an acceleration phase. 

The work developed in this section is the central result of our paper. 
Firstly, it shows that the perturbative approach described
in Ref.\cite{Barton2010b} strongly depends on the acceleration phase that
brings the atom to a constant velocity $v$. Secondly,
it proves that the description given in Ref.\cite{Barton2010b} of the quantum
friction is incomplete and that, when corrected, it is in agreement with
a drag force at zero temperature proportional to $v^{3}$.

%%%%%%%%%%

\section{Results from fluctuation electrodynamics}
\label{s:fluctQED}

In the previous sections we provided a complete description of quantum friction 
within the framework of perturbation theory. This approach has the merit of
relying on well-established techniques, even if the mathematical machinery
is somewhat cumbersome. Quantum friction, however, has been examined
within other frameworks, approaching the problem from other perspectives. 
For the sake of completeness, we review in this section some of the results from
fluctuation electrodynamics, which is one of the most used approaches to
describe the quantum mechanical interaction
of two neutral objects.

\subsection{Spectral densities}

Correlation functions of the atom and field variables are a convenient
way to characterize the atom-field interaction in terms of 
`resonant' and `non-resonant' processes. We start by collecting a
few formulas for the free observables and evaluate their correlations 
in the `bare' ground state denoted as $| g, {\rm vac} \rangle$. 

\paragraph{Field correlations.}
We use `time-ordered' correlations as they often appear in 
time-dependent perturbation theory. For the free scalar potential
and $t > t'$
\begin{eqnarray}
&& C_\Phi( \vec{r}, \vec{r}\,'\!, t - t' ) =
\langle {\rm vac} | 
\hat{\Phi}( \vec{r}, t ) \hat{\Phi}( \vec{r}\,'\!, t' ) 
| {\rm vac} \rangle
\label{eq:def-field-correlation}
\nonumber\\
&&=
\frac{ \hbar }{ 2\pi^2 }
\int\!\frac{ {\rm d}^2 k }{ k } \!
\int\limits_0^{\infty}\!{\rm d}\omega\,
\mathop{\rm Im}R( \omega )
{\rm e}^{ {\rm i} {\bf k} \cdot ( {\bf r} - {\bf r}' ) 
	- k ( z + z' ) }
{\rm e}^{ - {\rm i} \omega (t - t') }~.
\label{eq:field-correlation-1}
\end{eqnarray}
Evaluating this for an atom with constant velocity ${\bf v}$, 
we get ${\bf r}( t ) - {\bf r}( t' ) = ( t - t' ){\bf v}$
and observe that the vacuum spectrum extends into negative
frequencies, of the order 
${\cal O}( - {\bf k} \cdot {\bf v} ) =
{\cal O}( - {v} / z )$. This estimate is based on the natural
momentum cutoff provided by the exponential 
${\rm e}^{ - k ( z + z' ) }$. The rest of the frequency dependence
is governed by the reflection amplitude $R( \omega )$: a peak
at the surface plasmon resonance $\omega = \omega_S$ with width $\Gamma$
and an algebraic decay $\sim 1/\omega^3$ in the UV. In the
time domain, these features translate into a correlation that
oscillates at $\omega_S$ with an exponential envelope of width
$1/\Gamma$, plus an algebraic long-term tail $\sim 1/(t - t')^2$
that arises from the `Ohmic' behavior $\mathop{\rm Im}R( \omega )
\sim \omega$ for $\omega \to 0$.
For the electric field, evaluated along an atomic path parallel
to the surface, we get similarly
($\omega' = \omega - {\bf k} \cdot {\bf v}$, frequency in the
co-moving frame)
\begin{eqnarray}
% C_{ij}( \vec{r}( t ), \vec{r}( t' ), t - t' ) = 
&&
\langle {\rm vac} | 
\hat{E}_i( \vec{r}(t), t ) \hat{E}_j( \vec{r}( t' ), t' ) 
| {\rm vac} \rangle
\nonumber\\
&& =
\frac{ \hbar }{ 2\pi^2 }
\int\!{\rm d}^2 k 
\frac{ k_i k_j^* \,{\rm e}^{ - 2 k z } }{ k } \!
\int\limits_{- {\bf k} \cdot {\bf v}}^{\infty}\!{\rm d}\omega'\,
\mathop{\rm Im}R( \omega' + {\bf k} \cdot {\bf v} )
{\rm e}^{ - {\rm i} \omega' (t - t') }~.
\label{eq:}
\end{eqnarray}
Note that for a more general trajectory, the correlations are not
stationary, and more involved spectral representations like Wigner
or wavelet transforms would be needed.
The response function of the free field is known as the Green
function (tensor). Standard linear response theory 
gives
\begin{eqnarray}
G_{ij}( \vec{r}, \vec{r}\,'\!, t - t' ) &=& 
\frac{ {\rm i} }{ \hbar } \Theta( t - t' )
\langle {\rm vac} | 
\big[ \hat{E}_i( \vec{r}, t ), \hat{E}_j( \vec{r}\,'\!, t' ) \big]
| {\rm vac} \rangle
%	\label{eq:def-Green-function}
\nonumber	\\
&=&
\frac{ {\rm i} }{ 2\pi^2 }
\Theta( t - t' )
\int\!{\rm d}^2 k 
\frac{ {\rm e}^{ - k ( z + z' ) } }{ k } 
\,{\rm e}^{ {\rm i} {\bf k} \cdot ( {\bf r} - {\bf r}' ) }
\nonumber\\
&& {} \times 
\int\limits_0^{\infty}\!{\rm d}\omega\,
\mathop{\rm Im}R( \omega )
\big[ 
	k_i k_j^* \, {\rm e}^{ - {\rm i} \omega (t - t') }
	-
	k_j k_i^* \, {\rm e}^{ {\rm i} \omega (t - t') }
\big]
\label{eq:def-Green-function}
%	\label{eq:expansion-Green-tensor}
\end{eqnarray}
with an obvious evaluation along the path of a moving atom. (
It can be checked that the last line of Eq.(\ref{eq:def-Green-function}) 
agrees with the solution of the Maxwell equations for a point dipole
in the non-retarded approximation.)

%%%%%%%%

\paragraph{Dipole correlations.}

The free dipole operator shows in the theory of Ref.\cite{Barton2010b} 
a sharp line. 
It is actually the specific challenge of this model that the line
broadening appears self-consistently at higher orders in perturbation
theory.
In the atomic ground state
\begin{equation}
\langle g | \hat{D}_i( t ) \hat{D}_j( t' ) | g \rangle^{(0)}
=
\delta_{ij} 
\frac{ \hbar \alpha \Omega }{ 2 }
\,{\rm e}^{ -{\rm i} \Omega ( t - t' ) }
\label{eq:dipole-correlation-0}
\end{equation}
after summing over the degenerate excited states $| \vec{\eta} \rangle$.
In a simple scheme where the states $| \vec{\eta} \rangle$ have 
decay rates $\gamma_{\vec{\eta}}$, this could be generalized to
\begin{equation}
\langle g | \hat{D}_i( t ) \hat{D}_j( t' ) | g \rangle
\approx
\frac{ \hbar \alpha \Omega }{ 2 }
\sum_{\vec{\eta}}
\eta_{i}\eta_{j} 
\,{\rm e}^{ - {\rm i} \Omega ( t - t' ) - \gamma_{\vec{\eta}}|t - t'| / 2}
	\label{eq:d-d-simple-damping}
\end{equation}
giving a Lorentzian spectrum:
\begin{eqnarray}
S_{ij}( \omega ) = 
\int\!{\rm d}\tau\, {\rm e}^{ {\rm i} \omega \tau }
\langle g | \hat{D}_i( t + \tau ) \hat{D}_j( t ) | g \rangle
	\label{eq:def-dipole-spectrum}
%	\nonumber\\
= 
\frac{ \hbar \alpha \Omega }{ 2 }
\sum_{\vec{\eta}}
\frac{ \gamma_{\vec{\eta}} \, \eta_{i}\eta_{j} }
{ ( \omega - \Omega )^2 + \gamma_{\vec{\eta}}^2 / 4 }~.
	\label{eq:Lorentzian-dipole-spectrum}
\end{eqnarray}
This is also the result of master
equation techniques in combination with the regression 
formula~\cite{MandelWolf}.
Fermi's Golden Rule yields for the
decay rates in front of a smooth metallic surface 
[Eq.(2.10, 2.11) in Ref.\cite{Barton2010b}]:
%\begin{equation}
$\gamma_{\vec{\eta}} = 
\eta_i q_{ij} \eta_j \,
\gamma$, 
%\,,\qquad
$\gamma = (\alpha \Omega) 
	/(4 z^3)
\mathop{\rm Im} R( \Omega )$,
%	\label{eq:Bartons-line-widths}
%\end{equation}
where $q_{ij}$ is a dimensionless diagonal tensor with elements
$q_{xx} = q_{yy} = 1/2$ and $q_{zz} = 1$. We recognize again that
$\mathop{\rm Im}R( \omega )$ gives the spectral density of the
plasmon field.

The atomic response is given by the polarizability tensor
whose lowest approximation in the spectral domain is
\begin{eqnarray}
\alpha^{(0)}_{ij}( \omega ) &=& 
\frac{ {\rm i} }{ \hbar }
\int\limits_0^{\infty}\!{\rm d}\tau \,
{\rm e}^{ {\rm i} \omega \tau }
\langle g |
\big[ \hat{D}_i( t + \tau ), \hat{D}_j( t ) \big] | g \rangle
	\label{eq:def-polarizability}
	\nonumber\\
&=&
\delta_{ij} \frac{ d^2 }{ \hbar }
\left(
\frac{ 1 }{ \Omega - \omega - {\rm i} \lambda }
+
\frac{ 1 }{ \Omega + \omega + {\rm i} \lambda }
\right)
=
\frac{ \delta_{ij} \alpha \Omega^2 }{ \Omega^2 - (\omega + {\rm i} \lambda)^2 }~,
	\label{eq:polarizability-0}
\end{eqnarray}
where $\lambda$ is a positive infinitesimal that shifts the
frequency into the upper half-plane. This has the same structure
as for an oscillator. A simple finite-damping generalization would
replace $\lambda$ by the actual line widths:
\begin{equation}
\alpha_{ij}( \omega ) \approx
\sum_{\vec{\eta}}
\frac{ \eta_{i}\eta_{j} 
\alpha \Omega^2 }{ \Omega^2 - (\omega + {\rm i} \gamma_{\vec{\eta}})^2 }~.
	\label{eq:alpha-with-simple-damping}
\end{equation}
(For a discussion of the imaginary part near the anti-resonant
peak $\Omega + \omega \approx 0$, see 
Refs.\cite{Milonni2004,Milonni2008}.)
We note that for infinitely narrow lines, the dipole correlation
functions in Eqs. (\ref{eq:dipole-correlation-0}, \ref{eq:polarizability-0}) 
satisfy the zero-temperature
fluctuation--dissipation (FD) relation~\cite{Callen1951,WeissBook}
\begin{equation}
S_{ij}( \omega ) = 
2 \hbar \Theta( \omega ) 
\mathop{\rm Im} \alpha_{ij}( \omega )~,
	\label{eq:FD-relation}
\end{equation}
where $S_{ij}( \omega )$ is the dipole correlation 
spectrum in Eq.(\ref{eq:def-dipole-spectrum}).
The FD relation is not satisfied, however,
by the line-broadened expressions presented  in
Eqs.~(\ref{eq:d-d-simple-damping},\ref{eq:alpha-with-simple-damping}): the dipole
spectrum does not vanish like
$\mathop{\rm Im} \alpha_{ij}( \omega ) \sim \omega$ near zero
frequency, and extends also into the negative frequency band. In general, 
however, these expressions are the result of approximations.
We recall that the FD relation is valid under relatively mild equilibrium
requirements, in particular it also holds when the dynamics of
the dipole operator is non-linear~\cite{WeissBook,Polevoi1975}.
For driven systems like in our case, generalizations 
of the FD relation in Eq.(\ref{eq:FD-relation}) \cite{Agarwal1972} 
involve additional
`source' terms~\cite{Intravaia2014} or correlations of
observables that are conjugate with respect to entropy (production) 
rather than the Hamiltonian~\cite{Seifert2010}.

%%%%%%%%%

\subsection{Macroscopic QED with Markov approximation}
\label{s:macro-QED}

Scheel and Buhmann derived the quantum friction force on an atom
of arbitrary internal state from the average Lorentz 
force \cite{Scheel2009}, which, in the non-retarded limit, is 
given in Eq. (\ref{eq:force-operator}) at the beginning of
this paper.
The time evolution of the electric field is obtained by formally integrating
the equation of motion for the bosonic operators $\hat{a}_\kappa^{\phantom\dag}$,
$\hat{a}_\kappa^\dag$ with the result that 
%
%\begin{equation}
$\vec{\hat{E}}(\vec{r},t) =
\vec{\hat{E}}{}^{\rm free}(\vec{r},t) + \vec{\hat{E}}{}^{(S)}(\vec{r},t)$
%\end{equation}
%
with the free field operator
\begin{equation}
\vec{\hat{E}}{}^{\rm free}(\vec{r},t) = \int\limits_0^\infty\!{\rm d}\omega\,
\vec{\hat{\cal E}}(\vec{r},\omega) \, {\rm e}^{-{\rm i}\omega t} 
+
\mbox{h.c.}~.
\label{eq:E-Free}
\end{equation}
When the atom is not externally driven, then we find in normal 
ordering and an initial vacuum state for the field,
$\langle \cdots \vec{\hat{\cal E}}(\vec{r},\omega) \rangle 
= \langle \vec{\hat{\cal E}}{}^\dag(\vec{r},\omega) \cdots \rangle
= 0$.
When evaluating the radiative force $\vec{F}(t)$ 
in normal order, it thus turns out that it is
entirely due to radiation reaction, i.e., the relevant electric field
is the source field emitted by the atom at previous times. 
This can be written with the field's Green 
function [Eq.(\ref{eq:def-Green-function})]:
%, \ref{eq:def-source-field}
%
\begin{equation}
\hat{E}_i^{(S)}( \vec{x}, t ) =
\int\!{\rm d}t' \,
G_{ij}( \vec{x}, \vec{r}( t' ), t - t' ) \hat{D}_j( t' )~,
\end{equation}
where %, we recall, 
$\vec{\hat{D}}(t')$ is the dipole operator.
The velocity-dependent force is due to the delay in the radiation
reaction field:
the atom acts as a source for the electric field at a
previous point on its trajectory $\vec{r}(t')$; the generated
field then causes a force at a later position
$\vec{r}(t)$ where the atom has moved to in the meantime. For an
atom moving normal to the surface, Doppler shifts of atomic
transition frequencies and line widths give rise to additional
velocity-dependent effects. At retarded distances, the R\"ontgen
coupling of the moving atom to the electromagnetic field needs to be
taken into account~\cite{Scheel2009}.

For our problem with short (non-retarded) distances to the surface,
the Green function in Eq.(\ref{eq:def-Green-function}) yields a natural 
split of the
source field into positive and negative frequency components $\vec{\hat{E}}{}^{(S)}(\vec{x},t) = 
\vec{\hat{\cal E}}{}^{(S)}( \vec{x}, t ) + {\rm h.c.}$, where
\begin{eqnarray}
\label{II.9.10}
%\vec{E}^{(S)}(\vec{x},t) &=& 
%\vec{\cal E}^{(S)}( \vec{x}, t ) + {\rm h.c.}
%\nonumber\\
\vec{\hat{\cal E}}{}^{(S)}( \vec{x}, t ) &=&
\frac{ {\rm i} }{ 2\pi^2 }
\int\!{\rm d}^2 k 
\frac{ \vec{k} }{ k } 
\, {\rm e}^{ - k ( x_3 + z(t') ) } 
\int\limits_0^{\infty}\!{\rm d}\omega\,
\mathop{\rm Im}R( \omega )
\nonumber\\
&& {} \times
\int\limits_{-\infty}^{t}\!{\rm d}t'\,
{\rm e}^{ {\rm i} {\bf k} \cdot ( {\bf x} - {\bf r}( t' ) ) 
	- {\rm i} \omega (t - t') }
( \vec{k}^* \cdot \vec{\hat{D}}( t' ) )~.
\end{eqnarray}
In the following, we evaluate this at the position 
$\vec{x} = \vec{r}( t )$ of the atom 
and assume that the latter is moving at constant velocity
${\bf v}$ parallel to the surface. 
Although this is not the most general trajectory, we will later
argue that within the Markov approximation used in this subsection 
the precise history of how the
particle achieves its terminal velocity does not matter. 
The average Lorentz force thus becomes
\begin{eqnarray}
\vec{F}(t)
&=& 
\langle \hat{D}_i( t ) \vec{\nabla} {\hat{\cal E}}^{(S)}_{i}( \vec{r}(t), t ) 
+ \mbox{h.c.}
\rangle
\nonumber\\
&=&
- \frac{ 1 }{ 2\pi^2 }
\int\!{\rm d}^2 k 
\frac{ \vec{k} }{ k } 
\,{\rm e}^{ - 2 k z } 
\int\limits_0^{\infty}\!{\rm d}\omega\,
\mathop{\rm Im}R( \omega )
\nonumber\\
&& {} \times
\int\limits_{-\infty}^{t}\!{\rm d}t'\,
{\rm e}^{ - {\rm i} ( \omega -  {\bf k} \cdot {\bf v} ) (t - t') }
k_i k_j^*
\langle \hat{D}_i( t ) \hat{D}_j( t' ) \rangle
+ \mbox{c.c.}~.
	\label{eq:rad-reaction-force}	
\end{eqnarray}
In Ref.\cite{Scheel2009}, this expression was expanded for small
${\bf v}$; for the ease of comparison with other approaches,
we defer this to a later stage [Eq.(\ref{eq:scheel2009-force}) below].

For weak atom-field
coupling, we may evaluate the dipole-dipole correlation function 
$\langle \hat{D}_i( t ) \hat{D}_j( t' ) \rangle$
using
the Markov approximation. This entails converting the equations of
motion for the atomic operators into an integral equation and taking
the slowly-varying operators out of the integral. The result is 
an effective solution to the
equations of motion involving only operators at a single time $t$.
Hence, all memories of previous quantum states have been lost. As
shown in Ref.~\cite{Buhmann2008a}, the Markov approximation may
become invalid, e.g., if an excited atom near-resonantly interacts
with a narrow resonance of the medium-assisted field
(which is not the case here).
For our case of an atom initially
prepared in its ground state, 
the upshot of this analysis is the following representation of the
dipole correlation function in terms of lowering operators
$\hat{A}_{g\vec{\eta}}$ between the atomic levels: 
\begin{eqnarray}
\langle g | \hat{D}_i( t + \tau ) \hat{D}_j( t ) | g \rangle =
%&=&
d^2 
\sum_{\vec{\eta}}
\eta_i \eta_j 
\langle g | \hat{A}_{g\vec{\eta}}( t + \tau ) \hat{A}_{g\vec{\eta}}^\dag( t ) | g \rangle
	\label{eq:flip-operator-expansion}
\\
\frac{ {\rm d} }{ {\rm d}\tau }
\langle g | \hat{A}_{g\vec{\eta}}( t + \tau ) \hat{A}_{g\vec{\eta}}^\dag( t ) | g \rangle
=
%&=&
\big( - {\rm i}\Omega - {\textstyle\frac12}\gamma_{\vec{\eta}}
\big)
\langle g | \hat{A}_{g\vec{\eta}}( t + \tau ) \hat{A}_{g\vec{\eta}}^\dag( t ) | g \rangle~,
\label{eq:Markov-regression-formula}
\end{eqnarray}
where the second line contains the atomic frequency $\Omega$ and the 
line width
of the $|g \rangle \leftrightarrow | \vec{\eta} \rangle$ transition.
This also yields
the correlation function given in Eq.(\ref{eq:d-d-simple-damping}),
using in the initial condition the closure relation 
$\sum_{\vec{\eta}} \langle g | \hat{A}_{g\vec{\eta}}( t ) 
\hat{A}_{g\vec{\eta}}^\dag( t ) | g \rangle = 1$ for a ground-state atom.
Coming back to the radiation reaction force [Eq. (\ref{eq:rad-reaction-force})],
at large times the integral over $t'$ evaluates to

\begin{equation}
\int\limits_{-\infty}^t\!{\rm d}t'
\,{\rm e}^{ -{\rm i}( \omega - {\bf k} \cdot {\bf v} ) (t-t') }
\langle g | \hat{D}_i(t) \hat{D}_j(t') | g \rangle
=
d^2
% \frac{ \hbar \alpha \Omega }{ 2 }
\sum_{\vec{\eta}}
\frac{ \eta_{i}\eta_{j} 
	}{ 
	{\rm i} ( \Omega + \omega - {\bf k} \cdot {\bf v} )
+ \gamma_{\vec{\eta}} }~.
\label{eq:dipole-correlation-Markov}
\end{equation}
Thus for $t\to\infty$ the lateral force is
\begin{eqnarray}
{\bf F}
&=&
- \frac{ \hbar \alpha \Omega }{ (2\pi)^2 }
\int\!{\rm d}^2 k 
\frac{ {\bf k} }{ k } 
\,{\rm e}^{ - 2 k z } 
\int\limits_0^{\infty}\!{\rm d}\omega\,
\mathop{\rm Im}R( \omega )
\sum_{\vec{\eta}}
\frac{ | \vec{\eta} \cdot \vec{k}|^2 \gamma_{\vec{\eta}}
	}{ 
	( \Omega + \omega' )^2
+ \gamma_{\vec{\eta}}^2 / 4 }~,
\label{ForceMarkov}
\end{eqnarray}
where $\omega' = \omega - {\bf k} \cdot {\bf v}$ 
is again the Doppler-shifted frequency. If this were evaluated
for an infinitesimal (and isotropic) linewidth, we would recover
the first-order force from Eq.(\ref{eq:correc-I-f}), exponentially
suppressed for small $v$. Following Ref.\cite{Scheel2009}, we keep
a finite linewidth, observe that the
lateral force vanishes for an atom at rest, and expand for small $v$:
\begin{eqnarray}
{\bf F}
&\approx& 
- \frac{ \hbar \alpha \Omega }{ 2\pi^2 }
\int\!{\rm d}^2 k 
\frac{ {\bf k} ({\bf k} \cdot {\bf v}) 
	 }{ k } 
	 \,{\rm e}^{ - 2 k z } 
\sum_{\vec{\eta}}
| \vec{\eta} \cdot \vec{k}|^2 \gamma_{\vec{\eta}}
\int\limits_0^{\infty}\!{\rm d}\omega\,
\frac{ \mathop{\rm Im}R( \omega )
	}{ 
	( \Omega + \omega )^3 }~.
	\label{eq:scheel2009-force}	
\end{eqnarray}
We recall that this result holds for an atom moving parallel to a surface 
at nonretarded distances. 
Note that due to the Markov approximation made,
no memory of previous times is retained in the evolution equation for
the atomic variables. In particular, this means that this result does
not depend how the atom is accelerated to its final velocity
${\bf v}$.
The line widths $\gamma_{\vec{\eta}}$
for the smooth metal surface of the present model
have been given in the previous section where their anisotropy was also discussed. 
Note that this and the friction force were incorrectly 
given in Ref.~\cite{Scheel2009} due to an
error in the averaging over excited states.
The corrected calculation can be found in Ref.~\cite{BuhmannBookII}.
The sum over the excited states, weighted with their line widths, leads
to 
$\sum_{\vec{\eta}} |\vec{\eta} \cdot \vec{k}|^2 \gamma_{\vec{\eta}} 
= \frac32 \gamma k^2$,
where $\gamma = \gamma_{z}$ is the line width parameter 
for a perpendicular dipole. 
%[Eq.(\ref{eq:Bartons-line-widths})].
%
The frequency integral in Eq.(\ref{eq:scheel2009-force}) can be performed
with a Wick rotation to the imaginary axis 
\begin{equation}
\int\limits_{0}^{\infty}\!{\rm d}\omega
\frac{ \mathop{\rm Im} R( \omega ) }{ (\Omega + \omega)^3 }
=
\Omega\int\limits_{0}^{\infty}\!{\rm d}\xi
\frac{ \Omega^2 - 3 \xi^2 }{ (\Omega^2 + \xi^2)^3 }
R( {\rm i} \xi )
\approx
\frac{ \pi \omega_p^2 }{ 4 \omega_S (\Omega + \omega_S)^3 }~.
\label{eq:}
\end{equation}
where the last expression was obtained for a narrow
surface plasmon resonance ($\Gamma \ll \omega_S$, see 
also Eq.(\ref{eq:freq-integral})).
Performing the $k$-integral, we finally get for the lateral force
\begin{eqnarray}
{\bf F} &=&
- \frac{ 3 \hbar \alpha \Omega }{ 16\pi }
\frac{ \omega_p^2 \gamma }{ \omega_S (\Omega + \omega_S)^3 }
\int\!{\rm d}^2 k \, k \,
{\bf k} ({\bf k} \cdot {\bf v})
\,{\rm e}^{ - 2 k z }
\nonumber\\
&=&
- {\bf v}
\frac{ 9 \hbar \alpha^2 \Omega^3 }{ 512 z^8 }
\frac{ \omega_p^4 \Gamma 
}{  \omega_S (\Omega + \omega_S)^3 (\Omega^2 - \omega_S^2)^2 }~,
	\label{eq:Stefan-Stefan}
\end{eqnarray}
which gives a frictional power $- {\bf F} \cdot {\bf v}$
which agrees with the value for $P_{B}$ given in
Eq.(\ref{eq:power-PB-2}) and
first derived in Ref.\cite{Barton2010b}.
The significance of this agreement remains unclear at the moment due
to the very different underlying assumptions. The calculation reviewed
in this subsection depends only on the final atomic velocity, the details of its
launching procedure being lost in the memory-less Markovian
behaviour due to the finite correlation time resulting from atomic
dissipation (spontaneous decay). On the other hand, the time-dependent
perturbation theory of Ref.~\cite{Barton2010b} is valid for small
times and so implicitly assumes an infinite correlation time. It hence
depends on the atomic acceleration trajectory, where the agreement
with the above result is found only for a very specific out of many
possible choices: sudden acceleration. For a more meaningful
comparison, a calculation along the lines of Ref.~\cite{Barton2010b}
could be performed for a dissipative system with a finite correlation
time, where at sufficiently large times one would expect the result to
also be independent of the acceleration stage. 

%%%%%%%%%%%%

\subsection{Non-equilibrium dipole correlations}
\label{s:non-eq-FD}

The approach followed by Intravaia, Behunin and Dalvit \cite{Intravaia2014} combines techniques of fluctuation and macroscopic electrodynamics. 
While the expression for the radiation force has the same structure
as Eq.(\ref{eq:rad-reaction-force}) above, 
the calculation of the dipole correlation function
is performed differently.
In the limit $t\to \infty$, the system becomes stationary and the correlation function 
\begin{equation}
 C_{ij}(t,t-\tau)=\langle \hat{D}_{i}(t)\hat{D}_{j}(t-\tau)\rangle \to C_{ij}(\tau;{\bf v})
\end{equation}
depends only on the time difference $\tau = t-t'$ and the final
velocity ${\bf v}$. (Corrections due to the acceleration stage drop
out at this point.)
In the previous expression the dipole operator
$\vec{\hat{D}}(t)$ contains the exact dynamics of the moving atomic dipole (all orders in perturbation theory), i.e. including the backaction from the field/matter. The mean value has to be evaluated with respect to $\hat{\rho}_{\rm NESS}=\lim_{t\to \infty}\hat{\rho}(t)$ which defines the (in general unknown) density matrix describing the non-equilibrium steady state (NESS). 
The latter obviously depends on the atom's velocity ${\bf v}$; which is
why we added the second argument ${\bf v}$ to the correlation function.

For a dipole operator with the structure
$\vec{\hat{D}}(t) = d \sum_{\vec{\eta}}
\vec{\eta} (\hat{A}_{g\vec{\eta}}(t)+\hat{A}^{\dag}_{g\vec{\eta}}(t))$, 
the matrix $C_{ij}(\tau;{\bf v})$ is symmetric, and since stationarity implies $C_{ij}(\tau;{\bf v})=C^{*}_{ij}(-\tau;{\bf v})$, the power spectrum
\begin{equation}
S_{ij}(\omega; \mathbf{v}) =\int_{-\infty}^{\infty}\!{\rm d}\tau
\,{\rm e}^{ {\rm i} \omega \tau} C_{ij}(\tau; \mathbf{v})
	\label{eq:def-dipole-spec-2}
\end{equation}
is symmetric and real. The frictional force can then be written as
\begin{eqnarray}
{\bf F}
&=&
- \frac{ 1 }{ 2 \pi^2 }
\int\!{\rm d}^2 k 
\frac{ {\bf k} }{ k } 
\,{\rm e}^{ - 2 k z } 
\int\limits_0^{\infty}\!{\rm d}\omega\,
\mathop{\rm Im}R( \omega )
k_{i}k_{j}^{*}S_{ij}({\bf k} \cdot {\bf v}-\omega; \mathbf{v})~.
\label{eq:sourcemoving}	
\end{eqnarray}
In order to evaluate the previous expression one needs to know  $S_{ij}(\omega;{\bf v})$, which is in general available only within a perturbative approach. 
(An exception is an isotropic oscillator atom for which 
the dipole power spectrum can be found exactly~\cite{Intravaia2014}.)
For the model atom of Fig.\ref{fig:sketch-setup}(right),
there is a Pauli algebra 
for each excited state $|\vec{\eta} \rangle$, spanned by 
the operator $\hat{\sigma}_{1} = \hat{A}_{g\vec{\eta}}+\hat{A}^{\dag}_{g\vec{\eta}}=|\vec{\eta}\rangle\langle g|+|g\rangle\langle \vec{\eta}|$,  together with $\hat{\sigma}_{2}=i(|g\rangle\langle \vec{\eta}|-|\vec{\eta}\rangle\langle g|)$ and  $\hat{\sigma}_{3}=|\vec{\eta}\rangle\langle \vec{\eta}|-|g\rangle\langle g|$.
With the atom+field coupling $\hat{V}(t)=-\hat{D}_{i}(t)\hat{E}_{i}(\vec{r}(t),t)$, 
we have the following nonlinear equation of motion in the Heisenberg picture
\begin{equation}
\ddot{\hat{\sigma}}_{1}(t) + \Omega^{2} \hat{\sigma}_{1}(t) = 
- \frac{ 2 d \Omega }{ \hbar }  \hat{\sigma}_{3}(t)  \,
\vec{\eta} \cdot \vec{\hat{E}}(\vec{r}(t),t).
\label{exactTSA}
\end{equation}
We focus our attention on the computation of the two-time correlation tensor  
$C_{ij}(t,t';{\bf v}) =  d^{2}\sum_{\vec{\eta}}\eta_{i}\eta_{j}\mathcal{C}(t,t';{\bf v})$ with $\mathcal{C}(t,t';{\bf v})=\langle \hat{\sigma}_1(t) \hat{\sigma}_1(t')\rangle$.
To the lowest order in $d$, it can be evaluated from the free evolution
of the dipole operator, resulting in
%\begin{equation}
$C^{(2)}_{ij}(t,t';{\bf v}) =
d^{2}\delta_{ij}
\,{\rm e}^{- {\rm i} \Omega (t-t')}$
%\label{eq:}
%\end{equation}
which is nothing but Eq.(\ref{eq:dipole-correlation-0}). This results,
however, in a frictional force that is exponentially suppressed 
in $1/v$ (see also the discussion after Eq.(\ref{ForceMarkov})). To get a force scaling as a power law in $v$, 
one needs to include second-order radiative corrections in
$\mathcal{C}(t,t';{\bf v})$. 
To this end we first insert in Eq.(\ref{exactTSA}) the formal solution for the dynamics of  
\begin{equation}
\hat{\sigma}_{3}(t) = \hat{\sigma}_{3}(-\infty) + \frac{2 d}{\hbar\Omega}
\int\limits_{-\infty}^{t}{\rm d}t_{1}\, 
\dot{\hat{\sigma}}_{1}(t_1) \, \vec{\eta} \cdot \vec{\hat{E}}(\vec{r}(t_{1}),t_{1}), 
\label{eq:}
\end{equation}
and then replace the exact field $\vec{\hat{E}}(\vec{r},t)$ by its free evolution $\vec{\hat{E}}{}^{\rm free}(\vec{r},t)$, given in Eq.(\ref{eq:E-Free}).
This leads to an equation of motion correct to the second order in atom-field coupling:
\begin{eqnarray}
\ddot{\hat{\sigma}}_{1}(t) + \frac{2d^{2}}{\hbar^2} 
\int\limits_{-\infty}^{t}{\rm d}t_{1} &&\eta_{i}\eta_{j}\{\hat{E}_{i}^{\rm free}(\vec{r}(t),t),\hat{E}_{j}^{\rm free}(\vec{r}(t_{1}),t_{1})\}\dot{\hat{\sigma}}_{1}(t_{1}) \nonumber\\
&&+ \Omega^{2} \hat{\sigma}_{1}(t) 
= - \frac{2 d \Omega}{\hbar} \hat{\sigma}_{3}(0) \vec{\eta}\cdot \vec{\hat{E}}{}^{\rm free}(\vec{r}(t),t).
\label{second}
\end{eqnarray} 
Multiplying this equation from the right by $\hat{\sigma}_1(t')$ and taking the expectation value over the initial state $|g,{\rm vac} \rangle$
(we recall that the bare initial state can be used here because corrections to the NESS are captured in perturbation theory),
we get
\begin{eqnarray}
\ddot{\mathcal{C}}(t,t';{\bf v})+\Omega^{2}\mathcal{C}(t,t';{\bf v})
&&+\int_{-\infty}^{t}{\rm d}t_1 \mu(t-t_1;{\bf v})\dot{\mathcal{C}}(t_1,t';{\bf v})\nonumber\\
&&=-\frac{2 d \Omega}{\hbar}
\langle \hat{\sigma}_{3}(0) \vec{\eta} \cdot \vec{\hat{E}}{}^{\rm free}(\vec{r}(t),t)
\hat{\sigma}_{1}(t')\rangle
\label{eqC}
\end{eqnarray}
where the $\mu(\tau;{\bf v})$ is the Fourier transform of
\begin{equation}
\mu(\omega;{\bf v})=\frac{2 d^{2}}{\pi\hbar}\int {\rm d}^{2}k\,\mathrm{sign}(\omega+{\bf k} \cdot {\bf v})\, \frac{|\vec{\eta}\cdot\vec{k}|^{2}}{k} \,{\rm e}^{ - 2 k z } \mathop{\rm Im}R(\omega+{\bf k} \cdot {\bf v}) .
\end{equation}
From Eq.(\ref{second}), we also get the stationary solution 
for the dipole operator, correct to second order:
\begin{equation}
\hspace*{-10mm}
\hat{\sigma}_{1}(t) = -\frac{2 d \Omega}{\hbar}\int_{-\infty}^{\infty}\frac{{\rm d}\omega}{2\pi}\int \frac{{\rm d}^{2}k}{(2\pi)^{2}}\frac{\hat{\sigma}_{3}(0)
\vec{\eta} \cdot \vec{\hat{E}}{}^{\rm free}(\mathbf{k}, \omega+{\bf k}\cdot{\bf v})}{\Omega^2[1-\Delta(\omega;{\bf v})]-\omega^{2}- {\rm i} \omega \gamma(\omega;{\bf v})}{\rm e}^{{\rm i} (\mathbf{k}\cdot\mathbf{r}-\omega t)} ,
\label{solsigma}
\end{equation}
where $\gamma(\omega;{\bf v})=\mu(\omega;{\bf v})/2$ and 
($\mathrm{P}$ denotes the principal value)
\begin{equation}
\Delta(\omega;{\bf v})=-\mathrm{P}\int_{0}^{\infty}\frac{{\rm d}\omega'}{\pi} \frac{\omega^{2}}{\Omega^{2}} \frac{\mu(\omega',{\bf v})}{\omega'^{2}-\omega^{2}}.
\end{equation}
are both even in $\omega$ and %
give the second-order atomic frequency shift and decay rate. 
(They depend also on the
transition dipole $\vec{\eta}$ and on velocity.)

Finally, inserting Eq.~(\ref{solsigma}) into Eq.~(\ref{eqC}) and Fourier
transforming the resulting expression, we can write the dipole
spectrum~(\ref{eq:def-dipole-spec-2}) to fourth order in the 
dipole coupling as 
\begin{eqnarray}
S_{ij}(\omega;{\bf v}) = 2 \hbar
\int {\rm d}^{2}k\,\theta(\omega+{\bf k} \cdot {\bf v}) 
 &&\mathop{\rm Im}R(\omega+{\bf k} \cdot {\bf v})\,{\rm e}^{ - 2 k z }  \nonumber\\
 &&\times\alpha_{in}(\omega;{\bf v}) \frac{k_{n}k^{*}_{m}}{k} \alpha_{mj}^{*}(\omega;{\bf v})
\label{TSA_S}
\end{eqnarray}
where we defined  
\begin{equation}
\alpha_{ij}(\omega;{\bf v})=\sum_{\vec{\eta}}\frac{\alpha\eta_{i}\eta_{j}\Omega^{2}}{\Omega^{2}(1-\Delta(\omega;{\bf v}))-\omega^{2}-{\rm i}  \omega\gamma(\omega;{\bf v})} .
\end{equation} 
Note that this velocity-dependent polarizability differs from the `simple damping' velocity-independent form given in Eq. (\ref{eq:alpha-with-simple-damping}), as it contains non-Markovian memory effects 
through the 
frequency-dependent 
shift $\Delta(\omega; {\bf v})$ and damping $\gamma(\omega;{\bf v})$.
Using the symmetry in $\omega$ of all involved functions, one can show 
that $S_{ij}(\omega; {\bf v}) $ is even in ${\bf v}$ and that for
small velocities, it satisfies the fluctuation-dissipation relation:
\begin{eqnarray}
S_{ij}(\omega; {\bf v}) = 
2\hbar
\theta(\omega) \mathop{\rm Im}\tilde{\alpha}_{ij}(\omega)+ {\cal O}(v^2)
\label{Sexpansion}
\end{eqnarray}
where the imaginary part of the `dressed' polarizability is
\begin{equation}
\mathop{\rm Im}\tilde{\alpha}_{ij}(\omega) = 
\int {\rm d}^{2}k\,
\mathop{\rm Im}R(\omega) \,{\rm e}^{ - 2 k z }
\alpha_{in}(\omega;0)  \frac{k_{n}k^{*}_{m}}{k} \alpha_{mj}^{*}(\omega;0)~.
	\label{eq:Im-alpha-dressed}
\end{equation}
Using this in Eq.~(\ref{eq:sourcemoving}), we obtain the quantum friction force
to fourth order in the coupling, namely
\begin{eqnarray}
{\bf F}
=
- \frac{ 2\hbar}{\pi^{2}}
\int\!{\rm d}^2 k 
\frac{ {\bf k} }{ k } 
\,{\rm e}^{ - 2 k z } 
\int\limits_0^{\infty}\!{\rm d}\omega\,
&&\mathop{\rm Im}R( \omega )
\nonumber\\
&&\times 
k_{i}k_{k}^{*}\theta({\bf k} \cdot {\bf v}-\omega)\mathop{\rm Im} \tilde{\alpha}_{ij}({\bf k} \cdot {\bf v}-\omega)~.
	\label{eq:noneq-final-force}
\end{eqnarray}
The key observation is that the step 
function $\theta({\bf k} \cdot {\bf v}-\omega)$ limits
the $\omega$-integral to the narrow spectral range
$0 < \omega < {\bf k} \cdot {\bf v}$ of the anomalous Doppler effect.
For small velocities, we can expand the frequency-dependent functions
$\mathop{\rm Im}R( \omega )$ 
and 
$\mathop{\rm Im} \tilde{\alpha}_{ij}(\omega)$ around $\omega = 0$.
Only the first derivatives contribute since both functions 
are odd in $\omega$. 
One obtains in this way
\begin{eqnarray}
{\bf F}
& \approx & - \frac{45 \hbar {\bf v} v^{2}}{64 \pi z^7} 
\mathop{\rm Im} \tilde{\alpha}'(0) \mathop{\rm Im}R'(0),
\label{friction-exact}
\end{eqnarray}
where $\tilde{\alpha}'(0)$ is the frequency derivative of 
the dressed atomic polarizability 
given in Eq.(\ref{eq:Im-alpha-dressed}), evaluated for 
an atom at rest
(${\bf v}=0$) at distance $z$ from the surface. (For the full distance dependence,
one has to perform the $k$-integral in Eq.(\ref{eq:Im-alpha-dressed}) to obtain
$\mathop{\rm Im}\tilde\alpha_{ij}( \omega ) \sim 1 / z^3$.)
Note that, in contrast to the prediction of the previous subsection, 
non-equilibrium fluctuation electrodynamics results in a $v^{3}$-dependence 
for quantum friction.

A few remarks are in order.
First, the next-order term proportional to $v^2$ in the expansion of the dipole power spectrum in Eq.(\ref{Sexpansion}) leads to corrections to the frictional force proportional to $v^5$. 
Second, we note that the result in Eq. (\ref{friction-exact}),  derived from the fourth-order expansion of the dipole-dipole correlation for the moving two-level atom, and valid in the low velocity limit,  coincides with the result of fluctuation electrodynamics in local equilibrium  \cite{Dedkov2002,Dedkov2012,Pieplow2013} when the corresponding perturbative expression for the polarizability is used, and differs from the frictional power $P_A$ in 
Barton's calculation by a factor of 5
{}[see Eq.(\ref{eq:Barton-PA-in-scaling-limit})]. 
Third, the same expression for the frictional force in Eq.(\ref{friction-exact}) is obtained for the moving atom treated as an isotropic harmonic oscillator, a case in which exact expressions for the  dipole-dipole correlation and a non-equilibrium fluctuation-dissipation relation are available \cite{Intravaia2014}. Finally, it is possible to show that a peculiar cancellation occurs in the computation of the fourth-order dipole-dipole correlator for an atom moving at constant velocity, which translates into an exact cancellation of
terms linear in $v$ in the frictional force~\cite{Intravaia15prep}.
%  (Intravaia, Behunin, and Dalvit, in preparation).

%%%%%%%%%%%%%%

\section{Conclusion}

In summary, we have shown that the calculation of atom-surface quantum 
friction in the formulation based on perturbation theory \cite{Barton2010b}
depends on how the atom is boosted from being initially at rest to a configuration 
in which it is moving at constant velocity parallel to the planar interface.
We pointed out a subtle cancellation between the one-photon and part of the two-photon dissipating power. As a result the leading order contribution
to the frictional power is quartic in velocity.
Also, an alternative calculation of the average radiation force
leads to the same conclusions, that is atom-surface quantum friction 
scales as $v^{3}$.

We have reviewed recent calculations (Scheel and Buhmann~\cite{Scheel2009} and Intravaia et al.~\cite{Intravaia2014}) that generalize fluctuation electrodynamics
for the computation of the atom-surface quantum friction in the non-equilibrium stationary state. They agree on the way the friction
force is determined by the fluctuation spectrum of the dipole alone
[Eq.(\ref{eq:rad-reaction-force})], but differ in evaluating that spectrum,
in particular in the low-frequency regime where the anomalous Doppler
shift~\cite{Ginzburg1996} arises ($\omega \sim v / z$). This leads
in one case~\cite{Scheel2009} to a friction force linear in $v$, and in the 
other~\cite{Intravaia2014} to a $v^3$ force. To validate the master equation 
techniques behind these approaches and to resolve this discrepancy, 
it would be 
very interesting to extend the time-dependent perturbation theory pursued
here and to calculate, for example, atom-field correlations in the stationary
state.

%%%%%%%%%%%%

\paragraph*{Acknowledgments.}

SYB acknowledges support by the Deutsche Forschungsgemeinschaft (grant BU
1803/3-1). VEM thanks University of Potsdam for its hospitality while this
research was completed.
Work at Los Alamos National Laboratory was carried out under the auspicies of
the LDRD program. FI acknowledges 
financial support from the European Union Marie Curie People program through the 
Career Integration Grant No. 631571. We also thank Gabriel Barton for illuminating discussions.

%%%%%%%%

\appendix

\section{Two-photon emission}
\label{a:power-Barton}

\subsection{Leading to a force ${\cal O}( v )$}
\label{a:power-Barton-order-v2}

The result for the frictional power in Ref.\cite{Barton2010b} 
that turns out to
scale like ${\cal O}( v^2 )$, arises from the
following integral (Barton's notation $P_B$, Eq.(4.11) of 
Ref.\cite{Barton2010b} with the missing prefactor from the erratum)
\begin{eqnarray}
P_B &=& 
\frac{ \hbar \alpha^2 \Omega^2  }{ (2 \pi)^3 }
\int\!{\rm d}\kappa_1 \, {\rm d}\kappa_2
\, {\rm e}^{ - 2 (k_1 + k_2) z }
\frac{ ( {\bf k}_1 \cdot {\bf k}_2 - k_1 k_2 )^2 }{ k_1 k_2 }
\nonumber
\\
&& {} \times 
(\omega_1 + \omega_2 )
\mathop{\rm Im} R( \omega_1 )
\mathop{\rm Im} R( \omega_2 )
\frac{ \delta( \Omega - \omega_1' ) 
( {\bf k}_2 \cdot {\bf v} )^2 
	}{ (\Omega + \omega_2 )^2 ( \Omega + \omega_2' )^2 }
\label{eq:Barton-power-PB}
\end{eqnarray}
where, deviating from Barton's notation,
the prime denotes the Doppler-shifted frequencies, e.g.:
$\omega_1' = \omega - {\bf k}_1 \cdot {\bf v}$. We expand to the
leading order in $v$ and approximate 
$\delta( \Omega - \omega_1' ) / ( \Omega + \omega_2' )^2
\approx \delta( \Omega - \omega_1 ) / ( \Omega + \omega_2 )^2$ 
in the second line (drop the primes). The integrals
over the wave vectors ${\bf k}_{1,2}$ are then elementary and give
$9 \pi^2 v^2 / (16 z^8)$ -- this frictional power is quadratic in
the atomic velocity $v$.
The $\delta$-function fixes one frequency
to $\omega_1 = \Omega$. The remaining frequency integral is evaluated
for a narrow surface plasmon resonance, $\Gamma \ll \omega_S$. This
gives
\begin{equation}
\int\limits_0^\infty\!{\rm d}\omega\,
\frac{ \mathop{\rm Im} R( \omega ) }{ ( \Omega + \omega )^3 }
\approx
\frac{ \pi \omega_p^2 }{ 4 \omega_S (\Omega + \omega_S)^3 }
+
\frac{ \omega_p^2 \Gamma }{ 4 \Omega \omega_S^4 }~.
\label{eq:freq-integral}
\end{equation}
Barton gives the first term, and the second arises from the low-frequency
limit of the surface plasmon spectral density. It contributes in particular
in the regime $\Omega \sim \Gamma \ll \omega_S$. Putting everything
together, Barton's approach yields
\begin{eqnarray}
P_B &\approx& 
\frac{ 9 }{ 128 }
\frac{ \hbar \alpha \Omega v^2 }{ z^5 }
\underbrace{
	\frac{ \alpha \Omega \mathop{\rm Im} R( \Omega )
	}{ 4 z^3 }
	}_{\mbox{$\gamma$}}
\left[
\frac{ \omega_p^2 }{ \omega_S (\Omega + \omega_S)^3 }
+
\frac{ \omega_p^2 \Gamma }{ \pi \Omega \omega_S^4 }
\right]
\label{eq:Barton-power-PB-1}
\end{eqnarray}
where we have marked the excited state decay rate in the short-distance
limit [Eq.(2.11) of Ref.\cite{Barton2010b}].

%%%%%%

\subsection{Leading to a force ${\cal O}( v^3 )$}
\label{a:power-Barton-order-v4}

Barton's result for the frictional power that turns out to
scale like ${\cal O}( v^4 )$, arises from the
following integral (Barton's notation $P_A$, Eq.(4.11) of Ref.\cite{Barton2010b})
\begin{eqnarray}
P_A &=& 
\frac{ \hbar \alpha^2 \Omega^4 \omega_p^4 \Gamma^2 }{ 16 \pi^3 }
% \int\!{\rm d}^2k_1 \, {\rm d}^2 k_2
\int\!{\rm d}\kappa_1 \, {\rm d}\kappa_2
\, {\rm e}^{ - 2 (k_1 + k_2) z }
\frac{ ( k_1 k_2 - {\bf k}_1 \cdot {\bf k}_2 )^2 }{ k_1 k_2 }
\nonumber
\\
&& {} \times 
\frac{ \omega_1 \omega_2 ( \omega_1 + \omega_2) }{ 
|F( \omega_1, \omega_S )|^2 |F( \omega_2, \omega_S )|^2 }
\frac{ \delta( \omega_1' + \omega_2') 
}{ ( \Omega + \omega_1' )^2
( \Omega + \omega_2' )^2 }
\label{eq:Barton-power-PA}
\end{eqnarray}
where the prime denotes the non-relativistic Doppler shift:
$\omega_1' = \omega - {\bf k}_1 \cdot {\bf v}$. 
Note that the $\delta$-function enforces energy conservation
in the frame comoving with atom: the pair of plasmons has zero 
energy there, $\omega_1' + \omega_2' = 0$. Since the frequencies 
$\omega_{1,2}$ in the laboratory frame are positive, this condition
can only be satisfied if the Doppler shift is anomalous, 
for example $\omega_1' < 0$. 
The same condition also explains the spectrum of 
Cherenkov radiation \cite{Ginzburg1996}.

To evaluate the integral~(\ref{eq:Barton-power-PA}), 
we assume that the Doppler shift is small
enough. More precisely, note that the exponential
factor ${\rm e}^{ - 2 (k_1 + k_2) z }$ provides a typical range 
${\cal O}( 1 / z )$ for the $k$-vectors. 
The Cherenkov condition $0 \le \omega_1 + \omega_2 =
( {\bf k}_1 + {\bf k}_2 ) \cdot {\bf v}$
then restricts $\omega_{1,2}$ to the range
${\cal O}( v / z )$ and the required approximation is
$|{\bf k} \cdot {\bf v}| = {\cal O}( v / z ) \ll \Omega, \omega_S$. 
The frequency integrals then give in the leading order
\begin{eqnarray}
P_A &\simeq& 
\frac{ \hbar \alpha^2 \omega_p^4 \Gamma^2 }{ 96 \pi^3 \omega_S^8 }
\int\limits_{\makebox[0mm][l]{%
\qquad\footnotesize
$({\bf k}_1 + {\bf k}_2) \cdot {\bf v} \ge 0$}}%
\!
{\rm d}^2k_1 \, {\rm d}^2 k_2
\, {\rm e}^{ - 2 (k_1 + k_2) z }
\frac{ ( k_1 k_2 - {\bf k}_1 \cdot {\bf k}_2 )^2 }{ k_1 k_2 }
[({\bf k}_1 + {\bf k}_2) \cdot {\bf v}]^4~.
\label{eq:Barton-PA-2}
\end{eqnarray}
The restriction on the integration domain can be lifted, multiplying
with $\frac12$,
since the integrand is even under the transformation
$(k_{1x}, k_{1y}, k_{2x}, k_{2y}) \mapsto 
(-k_{2x}, k_{2y}, -k_{1x}, k_{1y})$ (${\bf v}$ points along the
$x$-axis). % (This was taken into account by Barton in the final result.)
The $k$-integrals then reduce to $27 \pi^2 / (16 z^{10})$, 
and we get Barton's Eq.(5.4)
\begin{equation}
P_A \simeq 
\frac{ 9 }{ 512 \pi }
\frac{ \hbar v^4 \alpha^2 \omega_p^4 \Gamma^2 }{ \omega_S^8 z^{10}}~.
\label{eq:Barton-PA-in-scaling-limit-a}
\end{equation}
Note that this expression cannot be written in terms of the 
(distance-dependent) decay rate %Eq.(\ref{eq:decay-rate})
which depends on the plasmonic mode density at the atomic resonance
$\Omega$. The calculation above illustrates that, on the contrary, 
the two-plasmon emission in $P_A$ is concentrated at much lower
frequencies ${\cal O}( v / z )$.
In the limit $\Omega \ll \omega_S$, however,
Eq.(\ref{eq:Barton-PA-in-scaling-limit-a})
contains exactly the same scaling 
compared to Eq.(\ref{friction-exact}) of the 
fluctuation electrodynamics, 
and is just smaller by a factor $1/5$.

Note that it is not obvious that the radiated power $P=P_{A}+P_{B}$
and the frictional power $- {\bf F} \cdot {\bf v}$ give the same
result, as the energy taken from the atomic motion may also be used
to excited the atom. This term, denoted by $d Q / dt$ by Dedkov
\& Kyasov~\cite{Kyasov2002relativistic}, is discussed in 
Secs.\ref{s:Barton-PA-1}, \ref{s:generic-path} for different
atomic trajectories.

%%%%%%%%%%%

\section{Piecewise constant acceleration}
\label{a:piecewise}

We evaluate in this appendix the state of the atom+field system 
in first and second order of perturbation theory, using an atomic
trajectory whose velocity increases continuously over a finite
time (thin blue in Fig.\ref{fig:paths}). This serves as a check
of the general result~(\ref{eq:asymptotic-split-A})  %ad(t) + B
in the long-time limit
and provides a complete list of terms that enter 
into the two-photon production rate~(\ref{eq:result-M}).

The one-photon amplitude $c_1^{(1)}( t )$ for the component
$| \vec{\eta}, \kappa \rangle$ of the state~(\ref{eq:state-expansion-FD}) 
is proportional to~(\ref{def:amplitude-c1-1})
\begin{equation}
\mathcal{A}( e,  {\bf k} \omega; t) 
= \int\limits_{-\infty}^{t} \!{\rm d}t_1 \; {\rm e}^{{\rm i} (\Omega+\omega) t_1} 
{\rm e}^{-{\rm i} {\bf k} \cdot {\bf r}(t_1)}~, 
\label{2nddef:amplitude-c1-1}
\end{equation}
where ${\bf r}(t_1)$ is the atomic path. This determines in the next
order the amplitude of 
$| \kappa_1, \kappa_2 \rangle$
to be proportional to
\begin{equation}
{\cal M}( {\bf k}_1\omega_1, {\bf k}_2 \omega_2; t ) =
\int\limits_{-\infty}^{t}\!{\rm d}t_2\,
\mathcal{A}( e,  {\bf k}_1 \omega_1; t_2 ) 
\, {\rm e}^{ {\rm i}(- \Omega + \omega_2) t_2 } 
\, {\rm e}^{ -{\rm i}{\bf k}_2 \cdot {\bf r}(t_2) } ~.
	\label{2ndeq:amplitude-c2}
\end{equation}
We consider a particular trajectory, namely a path with piecewise constant acceleration (see caption of Fig.\ref{fig:paths}):
\begin{equation}
{\bf r}( t ) = \left\{
\begin{array}{ll}
0 & t \le -\tau
\\[1ex]
\displaystyle 
\frac{ {\bf v} }{ 4 \tau } ( t + \tau)^2
& -\tau < t \le \tau
\\[1ex]
{\bf v} t
& \tau < t
\end{array}
\right.
\label{eq:}
\end{equation}
We define  $\beta = {\bf k} \cdot {\bf v}\tau$, and consider the limiting case of small velocity $\beta \ll 1$ and  `smooth launch' $|\Omega \pm \omega| \tau \gg 1$.
For simplicity, we condense the notation into
$\mathcal{A}( e,  {\bf k} \omega; t) \mapsto
\mathcal{A}_{e}( t )$,  
$\Omega + \omega \mapsto \omega$, 
and set $\tau \mapsto 1$.

\paragraph{One-photon process.}
To perform the integration Eq.(\ref{2nddef:amplitude-c1-1}), we consider
first the case that $t \le -1$. The integral is elementary:
\begin{equation}
t \le -1: \qquad
%\mathcal{A}( e,  {\bf k} \omega; t) 
\mathcal{A}_{e}( t )
= \frac{ {\rm e}^{{\rm i} \omega t} }{  {\rm i} \omega }
	\label{eq:Ae-before}
\end{equation}
assuming that the coupling is switched off at the lower limit.
For $- 1 < t \le 1$, the $t_1$-integral is split into
$- \infty < t_1 \le -1$, giving $\mathcal{A}_{e}( -1 )$ from
Eq.(\ref{eq:Ae-before}), 
and into $-1 < t_1 \le t$, which makes a phase factor appear
under the integral:
${\rm e}^{-{\rm i} {\bf k} \cdot {\bf r}(t_1)} = 
{\rm e}^{ - {\rm i} \beta ( t_1 + 1 )^2 / 4 }$. Since this
phase is $\le \beta \ll 1$, we expand this exponential and find
\begin{eqnarray}
\hspace*{-20mm}
- 1 < t \le 1: \ 
{\cal A}_e( t ) 
%&=& 
%{\cal A}_e( -1 ) + 
%\int\limits_{-1}^{t}\!{\rm d}x
%\; {\rm e}^{{\rm i} \omega x} 
%{\rm e}^{-{\rm i} \beta (x + 1)^2 / 4 }
%\nonumber\\
&\approx&
{\cal A}_e( -1 ) + 
\int\limits_{-1}^{t}\!{\rm d}x
\; {\rm e}^{{\rm i} \omega x} 
\left(
1 - \frac{ {\rm i} \beta }{ 4 } (x + 1)^2
+ {\cal O}( \beta^2 ) 
\right)
\nonumber\\
&=&
%{\cal A}_e( -1 ) + 
%\frac{ {\rm e}^{{\rm i} \omega t}  
%-
%{\rm e}^{- {\rm i} \omega} 
%	}{  
%	{\rm i} \omega }
\frac{ {\rm e}^{{\rm i} \omega t}  
	}{  
	{\rm i} \omega }
- 
%\frac{ {\rm i} \beta }{ 4 }
%\int\limits_{-1}^{t}\!{\rm d}x
%\; {\rm e}^{{\rm i} \omega x} 
%(x + 1)^2 
\frac{  \beta }{ 4 }
\frac{ {\rm e}^{{\rm i} \omega t } 
%	(t \omega +\omega -(1-i))
%	(t \omega +\omega +(1+i))
	( \omega^2 ( t + 1 )^2 + 
	2 {\rm i} \omega ( t + 1) 
	- 2 )
   + 2 {\rm e}^{-{\rm i} \omega }
   }{\omega ^3}
+ \ldots
	\label{eq:t2-integrated}
\end{eqnarray}
% The second expression vanishes for $t = -1$. 
When the acceleration is finished, we thus get
\begin{equation}
{\cal A}_e( 1 ) = 
\frac{ {\rm e}^{{\rm i} \omega}  
	}{  
	{\rm i} \omega }
- 
%\frac{ {\rm i} \beta }{ 4 }
%\int\limits_{-1}^{t}\!{\rm d}x
%\; {\rm e}^{{\rm i} \omega x} 
%(x + 1)^2 
\beta
\frac{ {\rm e}^{{\rm i} \omega } 
%	(t \omega +\omega -(1-i))
%	(t \omega +\omega +(1+i))
	}{\omega}
-
{\rm i} \beta
\frac{ {\rm e}^{{\rm i} \omega } 
%	(t \omega +\omega -(1-i))
%	(t \omega +\omega +(1+i))
	}{ \omega^2 }
+ \frac{ \beta ({\rm e}^{{\rm i} \omega }
- {\rm e}^{-{\rm i} \omega } ) }{ 2 \omega^3 }~.
\label{eq:}
\end{equation}
Finally, for larger times, the branch of the path with a constant
velocity contributes between $1 < t_1 \le t$, again an elementary
integral:
\begin{eqnarray}
t > 1:\
{\cal A}_e( t ) &=& 
{\cal A}_e( 1 ) +
 \frac{ {\rm e}^{{\rm i} ( \omega - \beta ) t} 
 - {\rm e}^{{\rm i} ( \omega - \beta )} 
	}{  {\rm i} ( \omega - \beta )}
\nonumber\\
&\approx&
{\cal A}_e( 1 ) +
\frac{ {\rm e}^{{\rm i} ( \omega - \beta ) t} 
	}{  {\rm i} ( \omega - \beta )}
- \frac{ {\rm e}^{{\rm i} \omega} 
	}{  {\rm i} \omega}
+ \beta \frac{ {\rm e}^{{\rm i} \omega} 
	}{  \omega}
+ {\rm i} \beta \frac{ {\rm e}^{{\rm i} \omega} 
	}{   \omega^2}~.
	\label{eq:Ae-middle}
\end{eqnarray}
We have expanded to first order in $\beta$ all terms except the one
where $\beta t$ appears in the exponent, because we shall be interested
in the long-time limit.
Note the three cancellations with ${\cal A}_e( 1 )$ so that we get
\begin{eqnarray}
t > 1:\
{\cal A}_e( t ) &\approx& 
\frac{ {\rm e}^{{\rm i} ( \omega - \beta ) t} 
	}{  {\rm i} ( \omega - \beta )}
+
\frac{ \beta ({\rm e}^{{\rm i} \omega }
- {\rm e}^{-{\rm i} \omega } ) }{ 2 \omega^3 }
=
\frac{ {\rm e}^{{\rm i} ( \omega - \beta ) t} 
	}{  {\rm i} ( \omega - \beta )}
+
\frac{ {\rm i} \beta }{ \omega^2 }
\frac{ \sin \omega }{ \omega }~.
	\label{eq:Ae-after}
\end{eqnarray}
Putting everything together and restoring the physical units, we get
\begin{eqnarray}
{\cal A}_e( {\bf k}_1\omega_1; t )  = 
\left\{ \begin{array}{ll}
\displaystyle
\frac{ {\rm e}^{{\rm i} (\Omega + \omega_1) t} }{  {\rm i} (\Omega + \omega_1) } & \mbox{for $t \le - \tau$} 
	\\[2ex]
\displaystyle
\frac{ {\rm e}^{{\rm i} ( \Omega + \omega_1 - \beta_1 ) t} }{  {\rm i} ( \Omega + \omega_1 - \beta_1 )}
+
\frac{ {\rm i} {\bf k}_1 \cdot {\bf v} }{ (\Omega + \omega_1)^2 }
\frac{ \sin (\Omega + \omega_1) \tau }{ (\Omega + \omega_1) \tau } & \mbox{for $t > \tau$}
\end{array}
\right.
\label{eq:A-late}
\end{eqnarray}
Note that the second line is exactly of the form put forward on general
grounds in Eq.(\ref{eq:asymptotic-split-A}) of the main text. The first
term corresponds to the `adiabatic limit' where the atomic velocity is
taken at its final value. It is independent of the duration $\tau$
of the acceleration. In the second term,
the sinc function (last fraction) reduces to unity
for a sudden acceleration (limit $\tau \to 0$). 
Any finite value of $\tau$ decreases this amplitude, and effectively
suppresses it when the atom is smoothly accelerated, i.e., 
$(\Omega + \omega_1) \tau \gg 1$.

\paragraph{Two-photon process.}
Its amplitude is given by integrating the one-photon amplitude
$\mathcal{A}( e, {\bf k}_1\omega_1; t_2 )$ once again 
{}[Eq.(\ref{2ndeq:amplitude-c2})]. We use the previous 
expression~(\ref{eq:A-late}) and Eq.(\ref{eq:Ae-middle}).
Two small parameters appear $\beta_{1,2} 
= {\bf k}_{1,2} \cdot {\bf v} \tau$ that we consider small and
of the same order. 

We begin for $t \le -1$ with an elementary integral
\begin{eqnarray}
t \le -1: \
{\cal M}( t ) &=& 
%\int\limits_{-\infty}^{t}\!{\rm d}t_2\,
%\frac{ {\rm e}^{{\rm i} (\Omega + \omega_1) t_2} 
%	}{  {\rm i} (\Omega + \omega_1) }
%\, {\rm e}^{ {\rm i}(- \Omega + \omega_2) t_2 } 
%\nonumber
%%
%\\
% &=&
-
\frac{ {\rm e}^{{\rm i} (\omega_1 + \omega_2) t} 
	}{ (\Omega + \omega_1) 
	(\omega_1 + \omega_2) }~.
\label{eq:}
\end{eqnarray}
For $-1 < t \le 1$, the first-order expansion in $\beta_1$, $\beta_2$
yields:
\begin{eqnarray}
\hspace*{-20mm}
-1 < t \le 1: \
{\cal M}( t ) &\approx& 
{\cal M}( -1 )
\nonumber\\
&& {} +
\int\limits_{-1}^{t}\!{\rm d}t_2\,
\frac{ {\rm e}^{{\rm i} (\omega_1 + \omega_2) t_2} 
	}{  {\rm i} (\Omega + \omega_1) }
\left(
1 - \frac{ {\rm i} \beta_2 }{ 4 } (t_2 + 1)^2
\right)
\nonumber
\\
&& {} 
- 
\beta_1 
\int\limits_{-1}^{t}\!{\rm d}t_2 \,
{\rm e}^{ {\rm i}(\omega_1 + \omega_2) t_2 } 
\left\{
\frac{  (t_2 + 1)^2 
	}{ 4 ( \Omega + \omega_1 )}
-
{\rm i} 
\frac{  (t_2 + 1)
	}{ 2 ( \Omega + \omega_1 )^2 }
+ \frac{ 1 }{ 2 ( \Omega + \omega_1 )^3 }
\right\}
\nonumber\\
&& {} +
\beta_1 \int\limits_{-1}^{t}\!{\rm d}t_2 \,
{\rm e}^{ {\rm i}(- \Omega + \omega_2) t_2 } 
\frac{ {\rm e}^{-{\rm i} ( \Omega + \omega_1 ) }  
}{ 2 ( \Omega + \omega_1 )^3 }
~,
\end{eqnarray}
where the last two lines arise from Eq.(\ref{eq:Ae-middle}).
The second line is an integral analogous to Eq.(\ref{eq:t2-integrated}),
and the result partially cancels with ${\cal M}( - 1 )$.
The other integrals are just a bit tedious to work out and eventually
yield the cumbersome expression
\begin{eqnarray}
{\cal M}(t) &\approx&
-
\frac{ {\rm e}^{{\rm i} (\omega_1 + \omega_2) t} 
	}{ (\Omega + \omega_1) 
	(\omega_1 + \omega_2) }
\nonumber\\
&& {} 
- \frac{ \beta_1 + \beta_2 }{ {\rm i} ( \Omega + \omega_1 ) }
{\rm e}^{{\rm i} (\omega_1 + \omega_2) t} 
\left\{
\frac{ (t + 1)^2 
	}{ 4 ( \omega_1 + \omega_2 )}
-
{\rm i} 
\frac{  t + 1
	}{ 2 ( \omega_1 + \omega_2 )^2 }
+ \frac{ 1 }{ 2 ( \omega_1 + \omega_2 )^3 }
\right\}
\nonumber\\
&& {} 
- \frac{ \beta_1 + \beta_2 }{ {\rm i} ( \Omega + \omega_1 ) }
\frac{ {\rm e}^{-{\rm i} (\omega_1 + \omega_2) } 
	}{ 2 ( \omega_1 + \omega_2 )^3 }
\nonumber\\
&& {} +
\frac{ {\rm i} \beta_1  }{ 2 ( \Omega + \omega_1 )^2 }
% Mathematica integral
\frac{ 
	{\rm e}^{{\rm i} ( \omega_1 + \omega_2 ) t } 
	( 1 - {\rm i}( \omega_1 + \omega_2 ) (t + 1) ) 
	- {\rm e}^{-{\rm i} ( \omega_1 + \omega_2 ) } 
	}{ ( \omega_1 + \omega_2 )^2 }
%\int\limits_{-1}^{t}\!{\rm d}t_2\,
%{\rm e}^{{\rm i} (\omega_1 + \omega_2) t} 
%(t_2 + 1)
\nonumber\\
&& {} 
- \frac{ \beta_1 }{ 2 ( \Omega + \omega_1 )^3 }
\frac{ 
{\rm e}^{{\rm i} (\omega_1 + \omega_2) t} 
-
{\rm e}^{- {\rm i} (\omega_1 + \omega_2)} 
	}{ {\rm i} (\omega_1 + \omega_2) }
\nonumber\\
&& {} 
+ \frac{ \beta_1 }{ 2 ( \Omega + \omega_1 )^3 }
\frac{ 
{\rm e}^{{\rm i} (-\Omega + \omega_2) t} 
{\rm e}^{- {\rm i} (\Omega + \omega_1)} 
-
{\rm e}^{- {\rm i} (\omega_1 + \omega_2)} 
	}{ {\rm i} (- \Omega + \omega_2) }~.
\label{eq:}
\end{eqnarray}
We finally get to the physically interesting case of late times
where Eq.(\ref{eq:Ae-after}) can be used and the integrals become
again elementary
\begin{eqnarray}
t > 1: {\cal M}( t ) &=& {\cal M}( 1 )
%+
%\int\limits_{1}^{t}\!{\rm d}t_2\,
%{\rm e}^{ {\rm i} ( - \Omega + \omega_2 - \beta_2 ) t_2 }
%\big\{
%\frac{ {\rm e}^{{\rm i} ( \Omega + \omega_1 - \beta_1 ) t_2 } 
%	}{  {\rm i} ( \Omega + \omega_1 - \beta_1 )}
%+
%\frac{ \beta_1 ({\rm e}^{{\rm i} ( \Omega + \omega_1 ) }
%- {\rm e}^{-{\rm i} ( \Omega + \omega_1 ) } ) }{ 2 ( \Omega + \omega_1 )^3 }
%\big\}
+
\int\limits_{1}^{t}\!{\rm d}t_2\,
\frac{ {\rm e}^{{\rm i} (  \omega_1 - \beta_1 + \omega_2 - \beta_2 ) t_2 } 
	}{  {\rm i} ( \Omega + \omega_1 - \beta_1 )}
\nonumber\\
&& {} +
\int\limits_{1}^{t}\!{\rm d}t_2\,
{\rm e}^{ {\rm i} ( - \Omega + \omega_2 - \beta_2 ) t_2 }
\frac{ {\rm i} \beta_1 \sin( \Omega + \omega_1 ) }{ ( \Omega + \omega_1 )^3 }
\nonumber\\
&=&
{\cal M}( 1 )
-
\frac{ {\rm e}^{{\rm i} ( \omega_1 - \beta_1 + \omega_2 - \beta_2 ) t } 
-
{\rm e}^{{\rm i} ( \omega_1 - \beta_1 + \omega_2 - \beta_2 ) } 
	}{ ( \Omega + \omega_1 - \beta_1 ) ( \omega_1 - \beta_1 + \omega_2 - \beta_2 )}
	\label{eq:M-after}\\
&& {} +
\frac{
{\rm e}^{ {\rm i} ( - \Omega + \omega_2 - \beta_2 ) t }
-
{\rm e}^{ {\rm i} ( - \Omega + \omega_2 - \beta_2 ) }
}{ - \Omega + \omega_2 - \beta_2 }
%\frac{ \beta_1 ({\rm e}^{{\rm i} ( \Omega + \omega_1 ) }
%- {\rm e}^{-{\rm i} ( \Omega + \omega_1 ) } ) 
%	}{ 2 {\rm i} ( \Omega + \omega_1 )^3 }
\frac{ \beta_1 \sin( \Omega + \omega_1 ) }{ ( \Omega + \omega_1 )^3 
}
	~.
\nonumber
\end{eqnarray}
This can be written as a sum ${\cal M}_A( t ) + {\cal M}_B( t )
+ {\cal M}_C$ whose terms we discuss separately now.

The `adiabatic amplitude' appears in the first line of Eq.(\ref{eq:M-after})
\begin{equation}
{\cal M}_A( t ) = -
\frac{ {\rm e}^{{\rm i} ( \omega_1' + \omega_2' ) t } 
	}{ ( \Omega + \omega_1' ) ( \omega_1' + \omega_2' )}
	\to
	-
\frac{ ( 2 \Omega + \omega_1' + \omega_2' )
	\,{\rm e}^{{\rm i} ( \omega_1' + \omega_2' ) t } 
	}{ ( \Omega + \omega_1' ) ( \Omega + \omega_2' ) ( \omega_1' + \omega_2' )}
\label{eq:}
\end{equation}
where the notation $\omega'_i = \omega_i - \beta_i$ was used. The
term after the arrow ($\to$) in this formula gives the total amplitude 
after symmetrizing the quantum numbers ${\bf k}_1\omega_1$ and 
${\bf k}_2\omega_2$ of the two plasmons.
This expression is independent of $\tau$ because
${\cal M}$ must be a squared time by definition. 
It is identical to the term featuring ${\rm e}^{ {\rm i} \Delta t }$
in Eq.(4.6) of Ref.\cite{Barton2010b}, the one that leads 
for $t \to \infty$
to the $\delta$-function $\delta( \omega_1' + \omega_2' )$
and the power $P_A$. 

The second line of Eq.(\ref{eq:M-after}) contains the other 
time-dependent term:
\begin{equation}
{\cal M}_B(t ) = 
\frac{ {\rm e}^{ {\rm i} ( - \Omega + \omega_2' ) t } 
}{ - \Omega + \omega_2' }
\frac{ ({\bf k}_1 \cdot {\bf v})  \sin ( \Omega + \omega_1 )\tau }
{ ( \Omega + \omega_1 )^3 \tau
	}~,
\label{eq:}
\end{equation}
where we have restored $\tau$.
To this order in ${\bf v}$, the limit $\tau \to 0$
recovers the term proportional to
${\rm e}^{ {\rm i} (- \Omega + \Delta_0') t }$
in Eq.(4.6) of Ref.\cite{Barton2010b}. 
We recall that this term leads to the 
$\delta$-function $\delta( \omega_2' - \Omega )$ and the power $P_B$
scaling with $v^2$, together with its exchange-symmetric partner.
Note that this term, up to the first factor, % $1/( - \Omega + \omega_2' )$,
is exactly given by the second (constant)
piece of the one-plasmon amplitude ${\cal A}_e( {\bf k}_1\omega_1; t)$ 
in Eq.(\ref{eq:A-late}). Hence the sinc function reduces its 
contribution if $(\Omega + \omega_1)\tau \gg 1$.

The remaining term ${\cal M}_C$ collects all terms independent of $t$
in Eq.(\ref{eq:M-after}).  Their expansion for small
$\beta_{1,2}$ is tedious and leads to
\begin{equation}
{\cal M}_C =
% symbolic algebra with Mathematica:
% \frac{i e^{-i (\omega_1+\omega_2)}
%   \left(-1+e^{2 i (\omega_1+\text{$\omega
%   $2})}\right) (\beta_1 (\Omega +\text{$\omega
%   $1})+\beta_2 (\Omega -\omega_2))
%   v}{2 (\Omega +\omega_1) (\text{$\omega
%   $2}-\Omega ) (\omega_1+\text{$\omega
%   $2})^3}+O\left(v^2\right)
%- \frac{
%   (\beta_1 + \beta_2 ) \Omega + \beta_1 \omega_1 - \beta_2 \omega_2
%	}{
%	(\Omega +\omega_1) (\omega_2 - \Omega ) (\omega_1 + \omega_2)^2}
%	\frac{ \sin (\omega_1 + \omega_2) \tau }{ (\omega_1 + \omega_2) \tau }
%	+ {\cal O}( \beta^2 )
\frac{ 1 }{ (\omega_1 + \omega_2)^2 }
\left\{
\frac{ \beta_1 }{
	\Omega - \omega_2 }
+
\frac{ \beta_2 }{ 
	\Omega + \omega_1 }
\right\}
	\frac{ \sin (\omega_1 + \omega_2) }{ (\omega_1 + \omega_2) }
	+ {\cal O}( \beta^2 )~.
\label{eq:}
\end{equation}
We add the corresponding expression under plasmon exchange 
($1 \leftrightarrow 2$) and get in physical units
\begin{equation}
{\cal M}_{C,{\rm sym}} =
%\frac{i e^{-i (\omega_1+\omega_2)}
%   \left(-1+e^{2 i (\omega_1+\text{$\omega
%   $2})}\right) v \Omega  \left(\beta_1
%   \left(\omega_1^2-\Omega
%   ^2\right)+\beta_2 \left(\text{$\omega
%   $2}^2-\Omega ^2\right)\right)}{\left(\Omega
%   ^2-\omega_1^2\right) (\text{$\omega
%   $1}+\omega_2)^3 \left(\Omega ^2-\text{$\omega
%   $2}^2\right)}
%-
%\frac{
%	2
%   \Omega  \left(\beta_1
%   \left(\omega_1^2-\Omega^2\right)+\beta_2 \left(\omega_2^2-\Omega ^2\right)\right)
%   }{
%   \left(\Omega^2 - \omega_1^2\right) \left(\Omega ^2 - \omega_2^2\right)
%   (\omega_1+\omega_2)^2 
%   }
%	\frac{ \sin (\omega_1 + \omega_2) \tau }{ (\omega_1 + \omega_2) \tau }
\frac{ 2 \Omega }{ (\omega_1 + \omega_2)^2 }
\left\{
\frac{ {\bf k}_1 \cdot {\bf v}
	}{ \Omega ^2 - \omega_2^2
   }
+
\frac{ {\bf k}_2 \cdot {\bf v}
	}{ \Omega^2 - \omega_1^2
   }
\right\}
	\frac{ \sin (\omega_1 + \omega_2) \tau }{ (\omega_1 + \omega_2) \tau }~.
\label{eq:MC-symmetrized}
\end{equation}
It is straightforward to check that this is equal to the small-${\bf v}$
expansion of the constant terms in ${\cal M}( t )$ as given 
in Eq.(4.6) of Ref.\cite{Barton2010b}.

To summarize, we have extended the 
calculation of the complete two-photon amplitude to an atomic path
with an acceleration phase of duration $2\tau$. The two-plasmon
power called $P_A$, scaling with ${\cal O}( v^4 )$ does not depend
on $\tau$ and is unchanged, at least for small velocities. 
The power $P_B$, scaling with ${\cal O}( v^2 )$, depends on $\tau$
and becomes suppressed when the launch duration $\tau$ is larger
than the atomic period $1/\Omega$. We have confirmed the argument given 
earlier that this ${\cal O}( v^2 )$ power can be computed from the
first-order transition amplitude: it is proportional to the
probability of exciting the atom in a non-adiabatic way. This means:
the acceleration has lead to an amplitude shift in the `Lamb cloud'
of virtual photons surrounding the atom. We may say that these photons
have become `real' because their amplitude differs from the adiabatic
value.

\paragraph{Technical note:} To get a probability amplitude that increases 
linearly with $t$, we need
\begin{equation}
\lim_{t \to \infty}
\frac{ {\rm e}^{ {\rm i} \nu t } - 1 }{ \nu }
= 2 \pi {\rm i} \delta( \nu )
\,,\qquad
\lim_{t \to \infty}
\bigg| \frac{ {\rm e}^{ {\rm i} \nu t } - 1 }{ \nu } \bigg|^2
= 2 \pi t \delta( \nu )~.
\label{eq:long-time-limit}
\end{equation}
If the `$-1$' is rather a complex function $a( \nu )$, we may evaluate
\begin{equation}
\bigg| \frac{ {\rm e}^{ {\rm i} \nu t } - 1 + 1 - a( \nu ) }{ \nu } \bigg|^2
\to 2 \pi t \delta( \nu )
%+ 2 \mathop{\rm Re} \frac{ ({\rm e}^{ {\rm i} \nu t } - 1)(1 - a^*)}{ \nu^2 }
+ 4 \pi \delta(\nu) \frac{ \mathop{\rm Im} a( \nu ) }{ \nu }
+ \frac{ |1 - a( \nu )|^2 }{ \nu^2 }
\label{eq:}
\end{equation}
where the last two terms do not increase with $t$ (if the final integral
converges, of course). Hence they drop out
when a transition rate is calculated.

%%%%%%%%%%%%%

\section{Evaluation of the average force}
\label{a:ave-force}

\subsection{Second order: exponentially small}

The second-order term of the average 
force~(\ref{eq:expansion-ave-force-3})
is given by
\begin{eqnarray}
{\bf F}^{(2)}( t ) &=&
2 \mathop{\rm Re} 
\sum_{\vec{\eta}} \!\int\!{\rm d}^3\kappa\,
\langle g,\mathrm{vac}| \mathbf{\hat{F}}(t) | \vec{\eta},\kappa \rangle
c_1^{(1)}( t )
	\label{eq:F1-force}\\
&=&
\frac{ 2 d^2 }{ \hbar } 
\mathop{\rm Re} 
\sum_{\vec{\eta}} \!
\int\!{\rm d}^3\kappa\,
{\bf k} |\vec{\eta} \cdot \vec{k}|^2 |\phi_{\kappa}|^2
\frac{ {\rm i} }{ 
( \Omega + \omega' - {\rm i} \lambda) }~,
\end{eqnarray}
where we have used the 
matrix element~(\ref{eq:ge-element-for-F})
and the amplitude 
$c_1^{(1)}( t )$ [Eq.(\ref{eq:c1-amplitude})].
We consider in this appendix only the long-time limit where
${\bf r}( t ) = {\bf v} t $. Summing over the excited states and
taking the real part, one gets:
\begin{eqnarray}
{\bf F}^{(2)}( t ) &=&
-
\frac{ 2 d^2 }{ \pi } 
\int\!{\rm d}^3\kappa\,
{\bf k} \, k
\, {\rm e}^{ - 2 k z }
\mathop{\rm Im} R( \omega )
\delta( \Omega + \omega' )
	\label{eq:F1-force-1}
\end{eqnarray}
which is nothing but Eq.(\ref{eq:correc-I-f}). The resonance 
condition $\Omega + \omega' = 0$ can only be satisfied for large
${\bf k} = {\cal O}( \Omega / v )$, making this contribution 
exponentially small in $v$.%

%%%%%%%%%

\subsection{Fourth order, via vacuum}

We continue with the fourth-order part involving the 
mixed amplitude $c_0^{(2)*}( t ) c_1^{(1)}( t )$ in
Eq.(\ref{eq:expansion-ave-force-3}). This product can be
combined with the last line of 
Eq.(\ref{eq:c3-adiabatically}) for $c_1^{(3)}( t )$ 
where we recognize the expression for $ c_1^{(1)}( t )$.
The sum yields a force
(subscript $0$ for `going via zero-photon sector')
\begin{eqnarray}
{\bf F}_0^{(4)}( t ) &=&
2 \mathop{\rm Re} 
\sum_{\vec{\eta}} \!\int\!{\rm d}^3\kappa\,
\langle g,\mathrm{vac}| \mathbf{\hat{F}}(t) | \vec{\eta},\kappa \rangle
c_1^{(1)}( t )
\nonumber\\
&& {} \times
\Big\{
- \gamma_g t - 
\frac{ {\rm i} \gamma_g / 2 - \delta E_g / \hbar }{
\Omega + \omega' - {\rm i} \lambda } 
\Big\}~.
	\label{eq:F3vac-1}
\end{eqnarray}
Here, we recognize one term, $- \gamma_g t \, {\bf F}^{(2)}( t )$,
quadratic in exponentially small parts, that translates the loss
of probability in the ground state. For the other piece, we use
the identity
\begin{eqnarray}
\hspace*{-10mm}
\mathop{\rm Re} \,{\bf k}
\frac{ \gamma_g / 2 + {\rm i} \delta E_g / \hbar }{
(\Omega + \omega' - {\rm i} \lambda)^2 } 
&=& 
\mathop{\rm Re}\Big\{ 
( \gamma_g / 2 + {\rm i} \delta E_g / \hbar )
\nabla_{\bf v}
\frac{ 1 }{
\Omega + \omega' - {\rm i} \lambda } 
\Big\}
\nonumber
\\
&=&
\frac{ \gamma_g }{ 2 }
\nabla_{\bf v}
\mathop{\rm Re}
\frac{ 1 }{
\Omega + \omega' - {\rm i} \lambda } 
- \frac{ \delta E_g }{ \hbar }
\nabla_{\bf v}
\mathop{\rm Im}
\frac{ 1 }{
\Omega + \omega' - {\rm i} \lambda } 
\label{eq:}
\end{eqnarray}
to identify ground-state decay rate and level shift from 
Eq.(\ref{eq:0-level-shift-and-rate})
\begin{eqnarray}
{\bf F}_0^{(4)}( t ) &=&
-
2 \mathop{\rm Re} 
\sum_{\vec{\eta}} \!\int\!{\rm d}^3\kappa\,
\langle g,\mathrm{vac}| \mathbf{\hat{F}}(t) | \vec{\eta},\kappa \rangle
c_1^{(1)}( t )
\frac{ {\rm i} \gamma_g / 2 - \delta E_g / \hbar }{
\Omega + \omega' - {\rm i} \lambda } 
\nonumber\\
&=& 
\frac{ 2 d^2 }{ \hbar } 
\mathop{\rm Re} 
\sum_{\vec{\eta}} \!
\int\!{\rm d}^3\kappa\,
{\bf k} |\vec{\eta} \cdot \vec{k}|^2 |\phi_{\kappa}|^2
\frac{ \gamma_g / 2 + {\rm i} \delta E_g / \hbar}{ 
( \Omega + \omega' - {\rm i} \lambda)^2 }
\nonumber\\
&=& 
\gamma_g
\nabla_{\bf v}
\frac{ d^2 }{ \hbar } 
\sum_{\vec{\eta}} \!
\int\!{\rm d}^3\kappa\,
|\vec{\eta} \cdot \vec{k}|^2 |\phi_{\kappa}|^2
\mathop{\rm Re} 
\frac{ 1 }{
\Omega + \omega' - {\rm i} \lambda }
\nonumber\\
&& \quad 
-
\frac{ \delta E_g }{ \hbar }
\nabla_{\bf v}
\frac{ 2 d^2 }{ \hbar } 
\sum_{\vec{\eta}} \!
\int\!{\rm d}^3\kappa\,
{\bf k} |\vec{\eta} \cdot \vec{k}|^2 |\phi_{\kappa}|^2
\mathop{\rm Im} 
\frac{ 1 }{
\Omega + \omega' - {\rm i} \lambda }
\nonumber\\
&=&
- \nabla_{\bf v}
\Big( \gamma_g \delta E_g \Big)~.
\label{eq:}
\end{eqnarray}
Again, this is an exponentially small term. For its interpretation,
one may think about the adiabatically stored energy in the
Lamb-shifted ground state $\delta E_g$.

%%%%%%%%

\subsection{Fourth order, via two photons}

The final piece for the force arises from that part of
$c_1^{(3)}( t )$ that goes via the two-photon sector
[first and second lines of Eq.(\ref{eq:c3-adiabatically})]. We have
to add the mixed term from the coherence between the one- and
two-photon sectors. The two contributions are denoted 
${\bf F}_2^{(4)[03]}$ and ${\bf F}_2^{(4)[12]}$ and are handled separately (subscript 2 for 'going via two-photon sector') :
\begin{eqnarray}
{\bf F}_2^{(4)[03]} &=&
2 \mathop{\rm Re} 
\sum_{\vec{\eta}} \!\int\!{\rm d}^3\kappa\,
\langle g,\mathrm{vac}| \mathbf{\hat{F}}(t) | \vec{\eta},\kappa \rangle
\bigg\{
\frac{ {\rm i}\, d^3 \phi_{\kappa}^* 
	\,{\rm e}^{ {\rm i} ( \Omega + \omega' ) t }
	}{ 
	\hbar^3 ( \Omega + \omega' - {\rm i} \lambda ) 
	}
\nonumber\\
&& \quad \times
\int\!{\rm d}^3\kappa_1\,
\frac{ (\vec{\eta} \cdot \vec{k}_1) |\phi_{\kappa_1}|^2 (\vec{k}_1 \cdot \vec{k})^* 
}{ 
( \omega_1' + \omega' - {\rm i} \lambda ) 
}
\Big(
\frac{ 1 }{ \Omega + \omega_1' - {\rm i} \lambda }
+
\frac{ 1 }{ \Omega + \omega' - {\rm i} \lambda }
\Big)
\bigg\}
\nonumber\\
&=& 
\frac{ 2 d^4 }{ \hbar^3 }
\mathop{\rm Re} 
\sum_{\vec{\eta}} \!\int\!{\rm d}^3\kappa\,{\rm d}^3\kappa_1\,
|\vec{k}_1 \cdot \vec{k}|^2 |\phi_{\kappa}|^2 |\phi_{\kappa_1}|^2
\frac{ {\rm i} {\bf k}
	}{ 
	( \Omega + \omega' - {\rm i} \lambda)
	( \omega_1' + \omega' - {\rm i} \lambda ) 
	}
\nonumber\\
&& \quad \times
\Big\{
\frac{ 1 }{ \Omega + \omega_1' - {\rm i} \lambda }
+
\frac{ 1 }{ \Omega + \omega' - {\rm i} \lambda }
\Big\}~.
\label{eq:}
\end{eqnarray}
To proceed, we neglect exponentially small terms arising from
the $\delta$-functions $\delta( \Omega + \omega' )$
and drop the $- {\rm i} \lambda$ in the corresponding non-resonant 
denominators. The only term that remains is (re-labeling
$\kappa \mapsto \kappa_2$ and Bose symmetrizing the integrand)
\begin{eqnarray}
{\bf F}_2^{(4)[03]} &\simeq&
-
\frac{ \pi d^4 }{ \hbar^3 }
\sum_{\vec{\eta}} \!\int\!{\rm d}^3\kappa_1\,{\rm d}^3\kappa_2\,
|\vec{k}_1 \cdot \vec{k}_2|^2 |\phi_{\kappa_1}|^2 |\phi_{\kappa_2}|^2
	\delta( \omega_1' + \omega_2' )
\nonumber\\
&& \quad \times
\Big\{
\frac{ {\bf k}_1
	}{ 
	\Omega + \omega_1'
	}
+
\frac{ {\bf k}_2
	}{ 
	\Omega + \omega_2'
	}
\Big\}
%\Big\{
%\frac{ 1 }{ \Omega + \omega_1' }
%+
%\frac{ 1 }{ \Omega + \omega_2' }
%\Big\}
\frac{ 2\Omega + \omega_1' + \omega_2' }{
	(\Omega + \omega_1')(\Omega + \omega_2') }~.
	\label{eq:F-2-03-integral}
\end{eqnarray}
Finally, we have to add the triple integral (symmetry factor $1/2$ 
cancels)
\begin{eqnarray}
{\bf F}_2^{(4)[12]} &=&
\mathop{\rm Re}
\sum_{\vec{\eta}} \!
\int\!{\rm d}^3\kappa \,{\rm d}^3\kappa_1 \,{\rm d}^3\kappa_2 \,
\langle \vec{\eta}, \kappa | \mathbf{\hat{F}}(t) 
	| g,\kappa_{1}\kappa_{2} \rangle 
c_1^{(1)*}( t )
c_2^{(2)}( t )~.
\end{eqnarray}
Insert the matrix element~(\ref{eq:eg2-element-for-F}) with its
two Bose-symmetric terms,  
the amplitudes~(\ref{eq:c1-amplitude}, \ref{eq:c2-amplitude}),
exploit the $\delta$-functions $\delta( \kappa - \kappa_{1,2} )$,
% and the sum rule~(\ref{eq:sum-rule}), 
to get
\begin{eqnarray}
{\bf F}_2^{(4)[12]} &=&
\frac{ d^4 }{ \hbar^3 }
\mathop{\rm Re}
\int\!{\rm d}^3\kappa_1 \,{\rm d}^3\kappa_2 \,
|\vec{k}_1 \cdot \vec{k}_2|^2 |\phi_{\kappa_1}|^2 |\phi_{\kappa_2}|^2
\nonumber\\
&& \times
\bigg\{
\frac{ {\rm i} }{ \omega_1' + \omega_2' - {\rm i}\lambda }
\Big(
\frac{ {\bf k}_2 }{ \Omega + \omega_1' + {\rm i}\lambda }
+
\frac{ {\bf k}_1 }{ \Omega + \omega_2' + {\rm i}\lambda }
\Big)
\nonumber\\
&& \times
\Big(
\frac{ 1 }{ \Omega + \omega_1' - {\rm i}\lambda }
+
\frac{ 1 }{ \Omega + \omega_2' - {\rm i}\lambda }
\Big)
\bigg\}~.
\end{eqnarray}
Neglecting again the imaginary part in the non-resonant denominators
we
observe that the same structure as Eq.(\ref{eq:F-2-03-integral})
emerges, up to a switch ${\bf k}_1 \leftrightarrow {\bf k}_2$ in 
the photon momenta. By symmetry, 
we expect that ${\bf F}$ and ${\bf v}$ are parallel and 
find for the projection
\begin{eqnarray}
{\bf v} \cdot ( {\bf F}_2^{(4)[03]} + {\bf F}_2^{(4)[12]})
&\simeq&
-
\frac{ \pi d^4 }{ \hbar^3 }
\sum_{\vec{\eta}} \!\int\!{\rm d}^3\kappa_1\,{\rm d}^3\kappa_2\,
|\vec{k}_1 \cdot \vec{k}_2|^2 |\phi_{\kappa_1}|^2 |\phi_{\kappa_2}|^2
	\delta( \omega_1' + \omega_2' )
\nonumber\\
&& \quad \times
{\bf v} \cdot ( {\bf k}_1 + {\bf k}_2 )
%\Big\{
%\frac{ 1 }{ \Omega + \omega_1' }
%+
%\frac{ 1 }{ \Omega + \omega_2' }
%\Big\}
\Big(
\frac{ 2\Omega + \omega_1' + \omega_2' }{
	(\Omega + \omega_1')(\Omega + \omega_2') }
\Big)^2~.
\label{eq:}
\end{eqnarray}
Under the $\delta( \omega'_1 + \omega'_2 )$, we may replace
${\bf v} \cdot ( {\bf k}_1 + {\bf k}_2 ) \mapsto
\omega_1 + \omega_2$, and we recover the structure of the 
two-photon emission
%, as written in Eq.(\ref{eq:Barton-PA})
\begin{eqnarray}
{\bf v} \cdot ( {\bf F}_2^{(4)[03]} + {\bf F}_2^{(4)[12]})
&\simeq&
-
\frac{ \pi d^4 }{ \hbar^3 }
\sum_{\vec{\eta}} \!\int\!{\rm d}^3\kappa_1\,{\rm d}^3\kappa_2\,
|\vec{k}_1 \cdot \vec{k}_2|^2 |\phi_{\kappa_1}|^2 |\phi_{\kappa_2}|^2
	\delta( \omega_1' + \omega_2' )
\nonumber\\
&& \quad \times
( \omega_1 + \omega_2 )
%\Big\{
%\frac{ 1 }{ \Omega + \omega_1' }
%+
%\frac{ 1 }{ \Omega + \omega_2' }
%\Big\}
\frac{ 4\Omega^2 }{
	(\Omega + \omega_1')^2(\Omega + \omega_2')^2 }
\nonumber\\
&=& - P_A~.
	\label{eq:PA-recovered}
\end{eqnarray}

%%%%%%%%%%%%%%%%%%%%%%%%%%%%%%%%%%%%%%%%%%%%%%%%%%%%%%%%%%%%%%%%%%%%%%%%%%%%%%%%

%====================================
%           Bibiliography           %
%====================================
%
% \newpage

\bigskip
%
%\newcommand{\bibpath}{../} 
%\newcommand{\bstpath}{} 
%
%% \bibliographystyle{vancouver}
%% \bibliographystyle{unsrt}
%\bibliographystyle{\bstpath iopart-num}
%%
%%\bibliography{qfr-literature}
%\bibliography{\bibpath qfr-literature}

\providecommand{\newblock}{}

\end{document}